\documentclass[%
superscriptaddress,
 amsmath,amssymb,
 aps,
 twocolumn,
 prb,
floatfix,
a4paper
]{revtex4-2}

\usepackage{amsfonts}
\usepackage{graphicx}
\usepackage[Symbol]{upgreek}
\usepackage{fancyhdr}
\fancypagestyle{plain}{
\fancyfoot[C]{\thepage}}
\renewcommand{\figurename}{Fig.}
\makeatletter
\renewcommand*{\fnum@figure}{{\normalfont\bfseries \figurename~\thefigure}}
\makeatother
\begin{document}

\title{Unravelling the Nature of Spin Excitations Disentangled from Charge Contributions in a Doped Cuprate Superconductor}

\author{Wenliang Zhang}
\email{wenliang.zhang@psi.ch}%
%\thanks{These authors contribute equally}
\affiliation{Photon Science Division, Swiss Light Source, Paul Scherrer Institut, CH-5232 Villigen PSI, Switzerland}
\author{Cli\`o Efthimia Agrapidis}
%\thanks{These authors contribute equally}
\affiliation{Institute of Theoretical Physics, Faculty of Physics, University of Warsaw, Pasteura 5, PL-02093 Warsaw, Poland}
\author{Yi Tseng}
\affiliation{Photon Science Division, Swiss Light Source, Paul Scherrer Institut, CH-5232 Villigen PSI, Switzerland}
\author{Teguh Citra Asmara}
\affiliation{Photon Science Division, Swiss Light Source, Paul Scherrer Institut, CH-5232 Villigen PSI, Switzerland}
\author{Eugenio Paris}
\affiliation{Photon Science Division, Swiss Light Source, Paul Scherrer Institut, CH-5232 Villigen PSI, Switzerland}
\author{Vladimir N. Strocov}
\affiliation{Photon Science Division, Swiss Light Source, Paul Scherrer Institut, CH-5232 Villigen PSI, Switzerland}
\author{Enrico Giannini}
\affiliation{Department of Quantum Matter Physics, University of Geneva, Quai Ernest-Ansermet 24, CH-1211 Gen\`eve 4, Switzerland}
\author{Satoshi Nishimoto}
\affiliation{Department of Physics, Technical University Dresden, 01069 Dresden, Germany}
\affiliation{Institute for Theoretical Solid State Physics, IFW Dresden, 01171 Dresden, Germany}
\author{Krzysztof Wohlfeld}
\email[]{krzysztof.wohlfeld@fuw.edu.pl}
\affiliation{Institute of Theoretical Physics, Faculty of Physics, University of Warsaw, Pasteura 5, PL-02093 Warsaw, Poland}
\author{Thorsten Schmitt}
\email[]{thorsten.schmitt@psi.ch}
\affiliation{Photon Science Division, Swiss Light Source, Paul Scherrer Institut, CH-5232 Villigen PSI, Switzerland}

\begin{abstract}

The nature of the spin excitations in superconducting cuprates is a key question toward a unified understanding of the cuprate physics from long-range antiferromagnetism to superconductivity. The intense spin excitations up to the over-doped regime revealed by resonant inelastic X-ray scattering bring new insights as well as questions like how to understand their persistence or their relation to the collective excitations in ordered magnets (magnons). Here, we study the evolution of the spin excitations upon hole-doping the superconducting cuprate Bi$_2$Sr$_2$CaCu$_2$O$_{8+\delta}$ by disentangling the spin from the charge excitations in the experimental cross section. We compare our experimental results against density matrix renormalization group calculations for a $t$-$J$-like model on a square lattice. Our results unambiguously confirm the persistence of the spin excitations, which are closely connected to the persistence of short-range magnetic correlations up to high doping. This suggests that the spin excitations in hole-doped cuprates are related to magnons---albeit short-ranged.

\end{abstract}
\maketitle

%\setlength{ indent}{12pt}
%\section*{INTRODUCTION}
\noindent{\bf INTRODUCTION}\\
Starting from antiferromagnetic Mott insulators, the cuprate high-temperature superconductors go through various quantum states with the charge carrier doping as the tuning parameter and form a universal doping-temperature phase diagram. For the hole-doped cuprates, superconductivity emerges in the intermediate doping regime in close proximity to the notorious pseudogap state and the recently established charge density wave state \cite{Keimer2015}. Disentangling the physics behind these intertwined states is a major challenge for constructing a complete theory of the superconducting cuprates. Fundamentally, the competition between the exchange energy of the localized spins and the kinetic energy of the doped holes is believed to dominate the basic physics, and the two energy scales are naturally regulated by the amount of doped holes \cite{Lee2006}. While the holes tend to delocalize and turn the system into a Fermi liquid at high doping level, electron correlations and local antiferromagnetic correlations survive with modest doping in the superconducting regime and coexist with the well-defined quasiparticles \cite{Lee2006,Xu2009,Damascelli2003,Tranquada2013}. Exactly how these spin and charge degrees of freedom act and interact throughout the doping-temperature phase diagram is therefore a crucial question towards the formulation of a definitive theory of superconductivity in the doped cuprates.

Experimentally, the dynamics and interactions of the spin and charge degrees of freedom are studied by assessing the momentum and energy dependence of the elementary excitations across the phase diagram. As suggested in numerous papers over the last 11 years \cite{Braicovich2010,LeTacon2011,Dean2013,Dean2013PRL,Dean2013PRB,LeTacon2013,Ishii2014,Lee2014,Dean2014,Guarise2014,Minola2015,Wakimoto2015,Peng2015,Ellis2015,Huang2016,Monney2016,Meyers2017,Ivashko2017,Minola2017,Chaix2018,Peng2018,Robarts2019}, resonant inelastic X-ray scattering (RIXS) observes intense spin excitations in a large momentum space area around the Brillouin zone center, which persist well into the over-doped region. Crucially the spin excitations in doped samples disperse along the ($\uppi$, 0) direction similarly to the magnons in the antiferromagnetic phase with the damping increasing moderately, although they rapidly become overdamped with doping along the ($\uppi$, $\uppi$) direction and show almost nondispersive profiles \cite{Dean2014,Guarise2014,Wakimoto2015,Huang2016,Monney2016,Meyers2017,Ivashko2017,Peng2018,Robarts2019}.
This suggests that in some particular area of the Brillouin zone, that is mostly in the ($\uppi$, 0) direction, the magnetic excitations completely `ignore' the existence of a critical value of the doping $\delta$ at which Fermi liquid behavior takes over the correlated magnetism and, moreover, that these excitations resemble the well-known magnons in undoped cuprates (hence their name---paramagnons). Naturally, such results are very much counterintuitive for they lead to an apparent paradox related to the small changes of the spin excitations in the doped cuprates, despite the rapid collapse of the long-range magnetic order upon doping and the dominant Fermi-liquid nature in the over-doped regime. This has sparked an intensive discussion on whether the observed magnetic excitations should indeed be viewed as paramagnons—or rather incoherent particle-hole excitations with a spin-flip \cite{Dean2014,Guarise2014,Minola2017,James2012,Benjamin2014,Nagy2016}.

To justify the nature of the spin excitations in cuprates as well as the reason of their persistence upon doping, it is necessary to precisely evaluate the momentum and doping evolution of the intrinsic spin excitations and comprehensively compare to theoretical calculations. This is a difficult task due to the experimental difficulties in extracting $S(\mathbf{q}, \omega)$ from RIXS spectra and also due to the problems in reliably calculating $S(\mathbf{q}, \omega)$. One of the major experimental obstacles comes from the mixing of spin and charge excitations in the RIXS spectra of doped cuprates \cite{Jia2016,Tsutsui2016}. With increasing doped holes, one would expect the charge excitations to be stronger, which will worsen the ‘mixing’ problem. This strongly hampers the correct assignment of the spectral profile to solely spin-flip containing excitations, and casts doubts on whether RIXS indeed observes the persistence of the intrinsic spin excitations.

Here we report a systematic study on the momentum and doping evolution of the disentangled intrinsic spin and charge excitations in superconducting Bi$_2$Sr$_2$CaCu$_2$O$_{8+\delta}$  (Bi2212). By applying the azimuthal-dependent analysis based on the distinct scattering tensors \cite{Kang2019}, we show that the low-energy excitations of the Cu $L_3$-edge RIXS spectra can be well described by spin-flip (spin) and non-spin-flip (charge) components for all studied doping levels and momenta, which allows us to extract the intrinsic spectral weights of the two components. We find that the spin spectral weight only slightly increases (decreases) with doping at intermediate momentum q along the ($\uppi$, $\uppi$) [($\uppi$, 0)] direction, which unequivocally confirms the persistence of the spin excitations in doped cuprates. We then compare the above experimental results to state-of-the-art density matrix renormalization group (DMRG) calculations of $t$-$J$-like models. The detailed comparison reveals the key characteristics of the spin excitations in the doped cuprates: On one hand, we unravel the crucial role of the longer-range hoppings, as the second- and third-nearest-neighbor hopping are needed to fully reproduce the experimental spin spectrum; on the other hand, we show that solely the short-range magnetic correlations are enough to qualitatively reproduce the persistence of the intensity of the spin excitations upon doping. Our results thus suggest that RIXS indeed observes the persistent spin excitations in hole-doped cuprates and that their paramagnetic nature can be understood as stemming from localised spins with short-range correlations.\\

\begin{figure}[t]
\centering
  \includegraphics[width=0.9\linewidth]{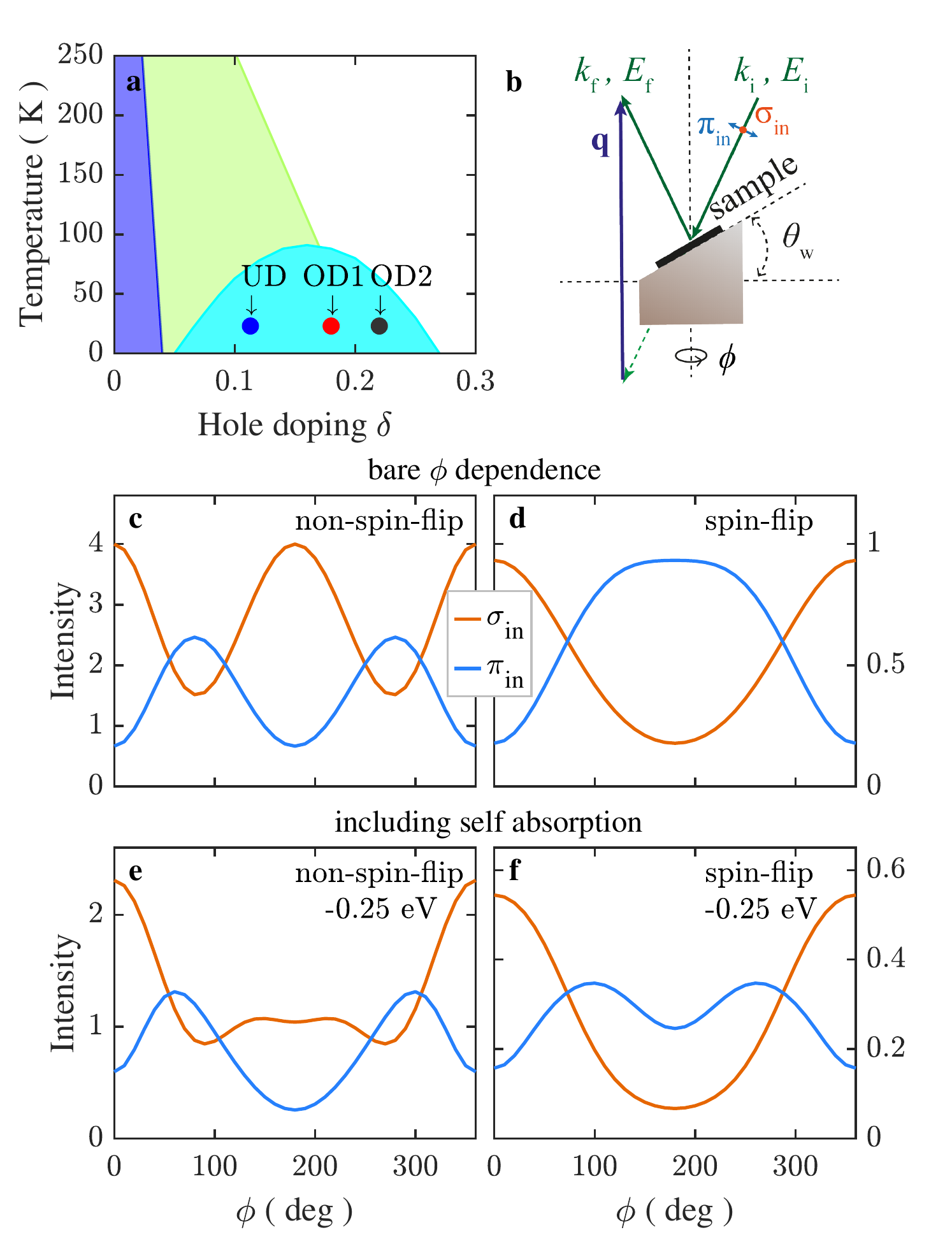}
  \caption{{\bf Experimental setup and the azimuthal dependence of non-spin-flip and spin-flip RIXS responses.} {\bf a} Schematic temperature-doping phase diagram for Bi2212. The purple, green and cyan areas represent the antiferromagnetic, pseudogap, and superconducting states, respectively. The solid circles indicate the locations of the three measured samples (UD, OD1, OD2) in the phase diagram. {\bf b} The scattering geometry of the azimuthal dependent RIXS experiment on a wedged sample. {\bf c} – {\bf f} The azimuthal $\phi$ dependence of the non-spin-flip and spin-flip RIXS responses in cuprates with a $40^\circ$ wedge angle ($\theta_\textrm{w}$). {\bf c} and {\bf d} are the ‘bare’ $\phi$ dependence on infinite thin samples, while {\bf e} and {\bf f} give the intensity evolution on bulk samples including the self-absorption effect at an energy loss of -0.25 eV.}
  \label{fig:boat1}
\end{figure}

%\section*{RESULTS}
\noindent{\bf RESULTS}\\
\noindent{\bf Disentangling the spin and charge excitations in the RIXS spectra}\\
%\subsection*{Disentangling the spin and charge excitations in the RIXS spectra}
We studied the spin and charge excitations of three different doped Bi2212 samples from under-doped to over-doped regime. The three samples are labeled as UD ($T_\textrm{c}$ = 73 K), OD1 ($T_\textrm{c}$ = 88 K)  and OD2 ($T_\textrm{c}$ = 65 K), as shown in Fig~\ref{fig:boat1}a. To disentangle the overlapping spin and charge excitations in the Cu $L_3$-edge RIXS spectra, we applied the recently proposed azimuthal dependent method \cite{Kang2019}, which resolves the two kinds of excitations based on their distinct scattering tensors. In this method, a sample on a wedged sample holder is rotated to change the orientation of the photon polarization in the sample space (see Fig.~\ref{fig:boat1}b and Methods section), which gives rise to different rotation dependences according to the scattering tensors \cite{Kang2019}. Fig.~\ref{fig:boat1}c,d show the azimuthal dependences of the non-spin flip and spin-flip excitations at $\sigma$ and $\pi$ incident polarizations with a 40$^\circ$ wedge angle. In the RIXS experiment, self-absorption effects need to be considered, which depends on the scattering geometry (azimuthal angles in this experiment) as well as the polarization and energy of the photons. Fig.~\ref{fig:boat1}e,f show the azimuthal dependence of spin and charge response after including the self-absorption effect at an energy loss ($E_\textrm{f} - E_\textrm{i}$) of -0.25 eV (see Supplementary Note 1). The spin and charge responses show clearly different azimuthal dependences, which allows to disentangle them in the low-energy RIXS spectra.

\begin{figure}[t]
\centering
  \includegraphics[width=0.9\linewidth]{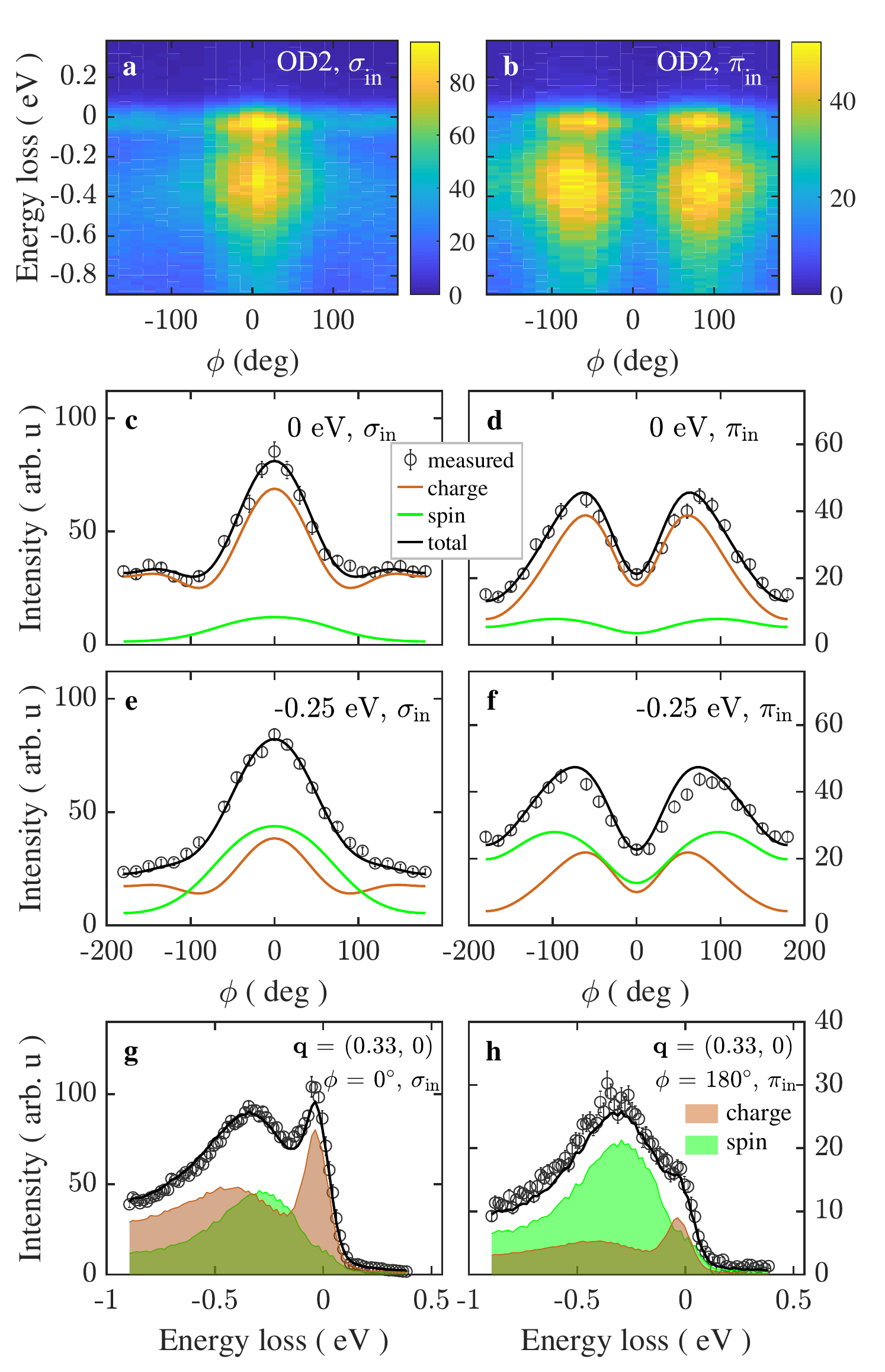}
  \caption{{\bf Azimuthal dependent RIXS spectra and the decomposition into the spin and charge contributions.} {\bf a} – {\bf b} Azimuthal dependence of the low-energy RIXS spectra of the OD2 sample, measured with $\sigma$ and $\pi$ incident polarizations, respectively, and at an in-plane momentum transfer of $\mathbf{q}$ = (0.33, 0). {\bf c} – {\bf f} Constant-energy-loss cuts of the azimuthal dependent RIXS intensity at 0 eV ({\bf c} and {\bf d}) and -0.25 eV ({\bf e} and {\bf f}). The solid green and red lines represent the decomposed spin and charge components $w_{\mathrm{s(c)}}{\cdot}A_{\mathrm{s(c)}}(\phi)$, respectively, and the solid dark lines are the sums of the two components. {\bf g} - {\bf h} The RIXS energy-loss spectra compared to the spin and charge components at grazing incidence ($\phi$ = 0$^\circ$) with $\sigma$ polarization and grazing emission ($\phi$ = 180$^\circ$) with $\pi$ polarization. The error bars represent the combined errors including the statistic errors and the errors in determining the zero-energy-loss positions (See Methods).}
  \label{fig:boat2}
\end{figure}

Fig.~\ref{fig:boat2}a and b show the RIXS intensity map of OD2 sample at $\mathbf{q}$ = (0.33, 0) as a function of azimuthal angle $\phi$ and energy loss $E$ at $\sigma$ and $\pi$ incident polarization, respectively. As the energies of $d$-$d$ excitations are situated well above 1 eV \cite{Dean2014,Guarise2014,Peng2017}, it is natural to assume that the low-energy excitations are mainly composed of the single-spin-flip and non-spin-flip excitations involving the 3$d_{x^2-y^2}$ orbital. Although the double spin flip with a net spin change of zero (bimagnon with $\Delta$$S$ = 0) is also allowed in RIXS process in cuprates \cite{Ament2011,Bisogni2012,Chaix2018}, earlier studies show that its spectral weight quickly diminishes with hole-doping \cite{Devereaux2007,Sugai2003,Sugai2013}, and becomes negligible in Cu $L_3$ RIXS when doping is beyond 0.08 \cite{Chaix2018}. In addition, the bimagnon intensity in Cu $L_3$ RIXS is maximal at the zone center, and becomes significantly weaker at large momentum \cite{Chaix2018}.
The low-energy RIXS intensity at a certain momentum $\mathbf{q}$ is thus expressed as a linear combination of the spectral weights of the single-spin-flip and non-spin-flip components modified by their corresponding azimuthal dependence:
\begin{equation}
    I_{\mathrm{RIXS}}(E, \phi, \boldsymbol{\epsilon}_\mathrm{i})=w_\mathrm{s}(E){\cdot}A_\mathrm{s}(E, \phi, \boldsymbol{\epsilon}_\mathrm{i})+w_\mathrm{c}(E){\cdot}A_\mathrm{c}(E, \phi, \boldsymbol{\epsilon}_\mathrm{i})
\label{eq:I_RIXS}
\end{equation}
where $w_{\mathrm{s(c)}}(E)$ is the spectral weight of spin (charge) components, and $A_{\mathrm{s(c)}}(E, \phi, \boldsymbol{\epsilon}_\mathrm{i})$ is the azimuthal dependence which is already known from the scattering tensor and self-absorption effects, and $\boldsymbol{\epsilon}_\mathrm{i}$ is the incident polarization. Fig.~\ref{fig:boat2}c-f show the $\phi$ dependence of RIXS intensity at constant energy loss of 0 and -0.25 eV with the decomposed spin and charge contributions $w_{\mathrm{s(c)}}{\cdot}A_{\mathrm{s(c)}}(\phi)$. As can be seen, the RIXS azimuthal dependences can be well fitted by these two components, which verifies that the above analysis correctly describes the low-energy RIXS response in Bi2212. In addition, the quasi-elastic scattering at 0 energy loss is dominated by charge-like $\phi$ dependence, consistent with the charge nature of the quasi-elastic peak. In Fig.~\ref{fig:boat2}g,h, we compare the RIXS spectra at two special geometries, grazing incidence ($\phi$ = 0$^\circ$) with $\sigma$ polarization and grazing emission ($\phi$ = 180$^\circ$) with $\pi$ polarization, with the decomposed spin and charge components. The grazing emission with $\pi$ polarization is usually used to measure the magnetic excitations in cuprates, since the charge component is largely suppressed in this geometry as shown in Fig.~\ref{fig:boat2}h. Nonetheless, the charge component is still considerable, which could influence the correct evaluation of the profile and intensity of magnetic excitations. It is therefore necessary to fully disentangle the spin and charge components to precisely study their nature. We note here that the obtained spectral functions $w_{\mathrm{s(c)}}(E)$ are solely related to the properties of the studied samples, as the angle and polarization related geometry factors and the self-absorption effect are removed by the knowledge of $A_{\mathrm{s(c)}}(E, \phi, \boldsymbol{\epsilon}_\mathrm{i})$. This allows the direct and unambiguous comparison between $w_{\mathrm{s(c)}}(E)$ and theoretical calculations based on different models, which could provide vital knowledge to understand the spin and charge excitations in cuprates.

\begin{figure*}[t]
  \includegraphics[width=0.9\textwidth]{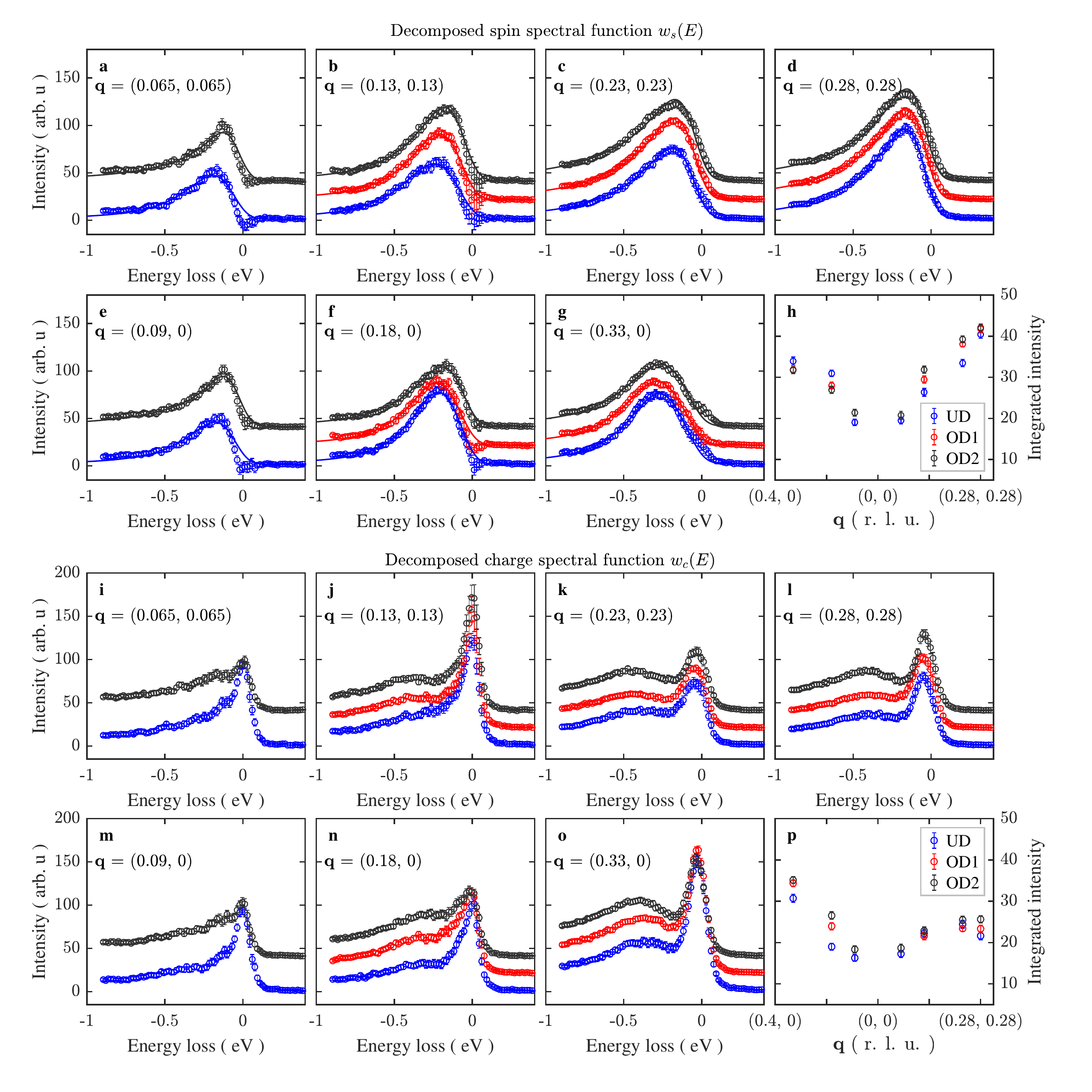}
  \caption{{\bf Momentum and doping dependence of the decomposed spin and charge spectral functions.} {\bf a} – {\bf g} ({\bf i} – {\bf o}) display the spin (charge) spectral functions for samples of different doping. The error bars indicate the combined errors of the fitting, statistic, and errors in determining the zero-energy-loss positions (See Methods). The solid lines in {\bf a} – {\bf g} indicate the fittings by a damped harmonic oscillator model. {\bf h} Energy integrated intensities of the spin spectral functions. {\bf p} Energy integrated intensities of the charge spectral functions with the quasi-elastic peaks subtracted. The error bars in {\bf h} and {\bf p} represent the integration errors assuming that the spectral errors have an energy correlation defined by the resolution function (see Methods).}
  \label{fig:boat3}
\end{figure*}

Fig.~\ref{fig:boat3} presents the obtained spin and charge spectral functions $w_{\mathrm{s(c)}}(E)$ of the three different doped samples at different momenta $\mathbf{q}$. Fig.~\ref{fig:boat3}a-g show the decomposed spin spectral functions, which all show a single peak with a damped profile. Error bars describe the fitting errors in the disentanglement (see Methods). Fig.~\ref{fig:boat3}h plots the energy integrated intensity $I_\mathrm{s}(\mathbf{q})$ of the spin spectral functions. The $I_s(\mathbf{q})$ of all three samples show a similar $\mathbf{q}$ dependence: $I_\mathrm{s}(\mathbf{q})$ monotonically increases with increasing $\mathbf{q}$, and has a slightly larger intensity when approaching large $\mathbf{q}$ along ($\uppi$, $\uppi$) direction than ($\uppi$, 0) direction. This $\mathbf{q}$ dependence is qualitatively consistent with the results in a previous study which calibrate the geometry influences by comparing to the INS results \cite{Robarts2019}.
Fig.~\ref{fig:boat3}i-o show the decomposed charge response at different $\mathbf{q}$.
There are two main components in the charge response: a quasi-elastic peak including the low-energy phonons close to zero energy loss, and a broad peak around -0.4 eV which extends to high energy loss. The quasi-elastic peak is enhanced at (0.13, 0.13), which is due to the structure modulation at (0.125, 0.125) in Bi2212 samples. Fig.~\ref{fig:boat3}p shows the integrated intensity of the broad peak after the quasi-elastic peak is subtracted.
In contrast to the spin response, this charge response shows much stronger intensity increases along ($\uppi$, 0) direction while it saturates around (0.25, 0.25) along ($\uppi$, $\uppi$) direction. The difference further highlights the distinct nature of the two decomposed components.

By comparing different doping levels, one can notice that the spin excitations show different development along the ($\uppi$, 0) and ($\uppi$, $\uppi$) directions: the total spectral weight increases with increasing doping at intermediate q along ($\uppi$, $\uppi$), while it slightly decreases along ($\uppi$, 0) direction, as shown by $I_\mathrm{s}(\mathbf{q})$ in Fig.~\ref{fig:boat3}h.
On the other hand, the spectral weights of the decomposed charge excitations simply increases with increasing doping along both directions, while the increase is more remarkable along ($\uppi$, 0) direction than ($\uppi$, $\uppi$) as shown in Fig.~\ref{fig:boat3}p.
A previous RIXS study on single layer (Bi,Pb)$_2$(Sr,La)$_2$CuO$_{6+\delta}$ \cite{Peng2018} also investigated the influence of doping on the spin excitations using the grazing-emission and incident $\pi$-polarization geometry which enhances the scattering contribution from the spin-flip channel. In contrast, they found the intensity of spin excitations increases with doping at small and intermediate $\mathbf{q}$ along both ($\uppi$, 0) and ($\uppi$, $\uppi$) directions, but crosses over to decrease at large $\mathbf{q}$. These different results could originate from either the differences between single and double layer cuprates, or a small residual mixture with charge excitations in the grazing-emission and incident $\pi$-polarization geometry. Note that the ($\uppi$, 0) direction will endure more influences from the residual charge excitations as the charge excitations are more intense and doping-dependent along the ($\uppi$, 0) direction (see Fig.~\ref{fig:boat3}p). The distinct spin excitations response to the hole doping along ($\uppi$, 0) and ($\uppi$, $\uppi$) directions could provide important information for the understanding of the spin dynamics in doped cuprates, which can be obtained by comparing with state-of-the-art theoretical calculations discussed in detail in the next part of the paper.

\begin{figure}[t]
\centering
  \includegraphics[width=\linewidth]{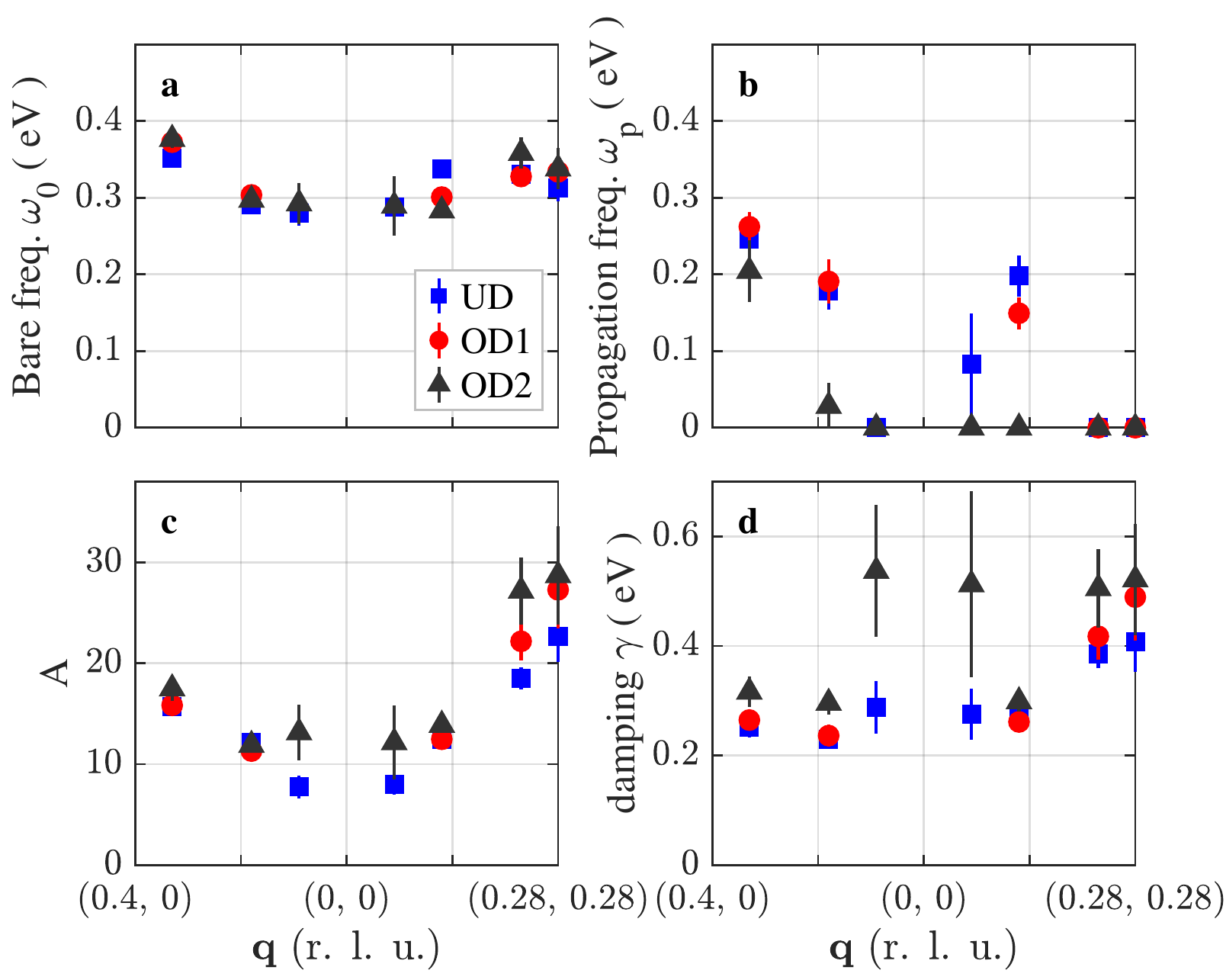}
  \caption{{\bf Fitting parameters of the spin spectral functions in Fig.~\ref{fig:boat3} by a damped harmonic oscillator model.} {\bf a} Bare frequency $\omega_0$. {\bf b} Propagation frequency $\omega_\mathrm{p}$. {\bf c} Amplitude $A$. {\bf d} Damping $\gamma$. The error bars represent the fitting errors.}
  \label{fig:boat4}
\end{figure}

For a more quantitative analysis of the spin response, we fit the spin spectral functions $w_{\mathrm{s(c)}}(E)$ by a damped harmonic oscillator (DHO) model convoluted with a resolution function (see Methods). As shown by the solid lines in Fig.~\ref{fig:boat3}a–g, the results can be well fitted by the DHO model. Fig.~\ref{fig:boat4} presents the fitting results. The bare frequency $\omega_0$ is similar for all three dopings, while the damping $\gamma$ increases with increasing doping and has a much larger value along ($\uppi$, $\uppi$) direction. This is consistent with previous studies on doped cuprates showing that the magnetic excitations are much more damped along ($\uppi$, $\uppi$) direction \cite{Dean2014,Guarise2014,Wakimoto2015,Huang2016,Monney2016,Meyers2017,Ivashko2017,Peng2018,Robarts2019}. With the charge excitations excluded, we can now rule out that the over-damped profile of the spin excitations along ($\uppi$, $\uppi$) direction comes from an increase of charge contributions with doping. We note that the fitted damping factors of the two smallest momenta of OD2 sample are large and out of the main trend of the momentum dependence. We attribute this to the decomposed shape of spin excitations bearing more influences from the uncertainties in the azimuthal fitting when the spin excitation peak is getting closer to the elastic peak at small momenta. Fig.~\ref{fig:boat4}b shows the propagation frequency defined as $\omega_\mathrm{p} = \sqrt{\omega_0^2 - \gamma^2}$, where a zero value is assigned when the system is over damped, i.e. $\omega_0 < \gamma$. One can see that the spin excitations in the over-doped OD2 sample are fully over damped along the ($\uppi$, $\uppi$) direction. Fig.~\ref{fig:boat4}c shows the fitted proportional amplitude $A$ to DHO model. It increases with increasing doping along ($\uppi$, $\uppi$) direction while it changes little along ($\uppi$, 0) direction, which is a bit different from the integrated total spectral weight $I_\mathrm{s}(\mathbf{q})$ shown in Fig.~\ref{fig:boat3}h. This is mostly due to $I_\mathrm{s}(\mathbf{q})$ including both the effects from the proportional amplitude $A$ and the damping $\gamma$, while $A$ excludes the effect of damping $\gamma$ which suppresses the total intensity.\\

%\subsection*{Unravelling the character of the spin excitations in doped cuprates}
\noindent{\bf Unravelling the character of the spin excitations in doped cuprates}\\
The model of choice to study the evolution of the spin excitations upon doping the cuprates is the ``celebrated'' $t$-$J$-like model \cite{Chao1978}, defined by the Hamiltonian on a 2D square lattice:
\begin{align}
H=&-t \sum_{\langle i,j \rangle} (\tilde c_i^\dagger\tilde c_j + h.c.) -t'\sum_{\langle\langle i,j\rangle\rangle}(\tilde c_i^\dagger\tilde c_j  + h.c.) \nonumber \\
&-t''\sum_{\langle\langle\langle i,j\rangle\rangle\rangle}(\tilde c_i^\dagger\tilde c_j+ h.c.)
+ J \sum_{\langle i,j \rangle} \left(\mathbf S_i \cdot \mathbf S_j  - \frac{1}{4} \tilde n_i\tilde n_j\right),
\end{align}
where $\tilde c_i^\dagger$ operator creates an electron at site $i$ in the constrained Hilbert space without double electron occupancies, $\mathbf S_i$  is a spin-1/2 operator at site $i$ and $\tilde n_i$  is the on-site electron density at site $i$: $\tilde n_i=\tilde c_i^\dagger \tilde c_i$ . The model parameters $t$, $t'$ and $t''$  denote the hopping integrals between first, second, and third neighbors, respectively, whereas $J$ is the antiferromagnetic Heisenberg interaction between nearest neighbor spins. In our calculations, we take a realistic and widely-accepted (cf. \cite{Dellanoy2009}  or a quite similar choice in ~\cite{Senechal2004}) choice of the values of the $t$-$t'$-$t''$-$J$ model parameters $t'=-0.3 t$, $t''=0.15 t$ and $J=0.4 t$. These values slightly differ from the doping-dependent values suggested for Bi2212 in \cite{Drozdov2018}, but we keep these more standard values to make our study universal. Furthermore, to better understand the role of the longer-range hoppings, we consider switching off the $t''$ hopping (the $t$-$t'$-$J$ model), both the $t''$ and $t'$ hoppings (the $t$-$J$ model) as well as substantially increasing the value of $t'$ in the $t$-$t'$-$J$ model calculations. Finally, note that, while to describe electronic properties of the cuprates probably the charge-transfer ($pd$) model would be more appropriate, the spin excitations are believed to be well-described by models with oxygens being integrated out \cite{Jia2014, Jia2016,Peng2018,Jia2014,Eder1995,Tohyama1995, Nocera2017,Tohyama2018,Parschke2019}. The $t$-($t'$-$t''$-)$J$ model on a square lattice has been intensively studied as a host of the superconducting state for a long time (for example, see \cite{Jiang2021,Gong2021} and references therein), and a possible existence of the superconducting phase in a wide range of hole doping has been proposed \cite{Jiang2018}.

Next our goal here is to compute the static spin structure factor $S(\mathbf{q})$, typically defined as:
\begin{align}
S(\mathbf{q})&= \langle \mathbf S(-\mathbf{q})\cdot \mathbf S(\mathbf{q})\rangle
=\frac{1}{N}\sum_{i,j} \langle \mathbf S_i \cdot \mathbf S_j \rangle \mathrm{e}^{i\mathbf{q}(\textbf{r}_i-\textbf{r}_j)}, \label{eq:FT}
\end{align}
where the $i$, $j$ indices run over all sites, $N$ is the number of sites and $\textbf{r}_i$ denotes the position of the site in the lattice. This is done by involving state-of-the-art DMRG calculations which are carried out on an $N=6\times6$ square lattice with open boundary conditions (OBC). We chose a 6x6 OBC cluster to investigate a wide range of parameters with high accuracy. The use of OBC enables us to avoid an artificial enhancement of specific period correlations, which frequently occurs in periodic and cylindrical systems. However, due to the OBC, the charges tend to localize at the edges when charge imbalance, i.e. holes, is introduced. To counterbalance this effect, we introduce an edge factor $\lambda$ that multiplies the electronic hopping parameters $t$, $t'$ and $t''$ as well as the spin exchange coupling $J$ acting on the sites on the perimeter of our cluster, see Methods and Supplementary Note 6.

By using the real space approach we have full control over which contributions to $S(\mathbf{q})$ we include in Equation \ref{eq:FT}. In fact, we can select the distance $\ell$ at which the sum in Equation \ref{eq:FT} is truncated. In particular, it is possible to consider the different total averages for the different spin-spin correlations $\langle \mathbf S \cdot \mathbf S\rangle_\ell $ with $\ell=1, 2, 3, \dots$ defining the considered neighbors. The difference in $S(\mathbf{q})$ between using the singular values of the spin-spin correlations and the averages is minimal (see Supplementary Note 7). While shielding us from accessing $S(\mathbf{q}, \omega)$, this approach allows us to thoroughly study the possible magnonic character of the persistent spin excitations.

\begin{figure}[t]
\centering
  \includegraphics[width=0.9\columnwidth]{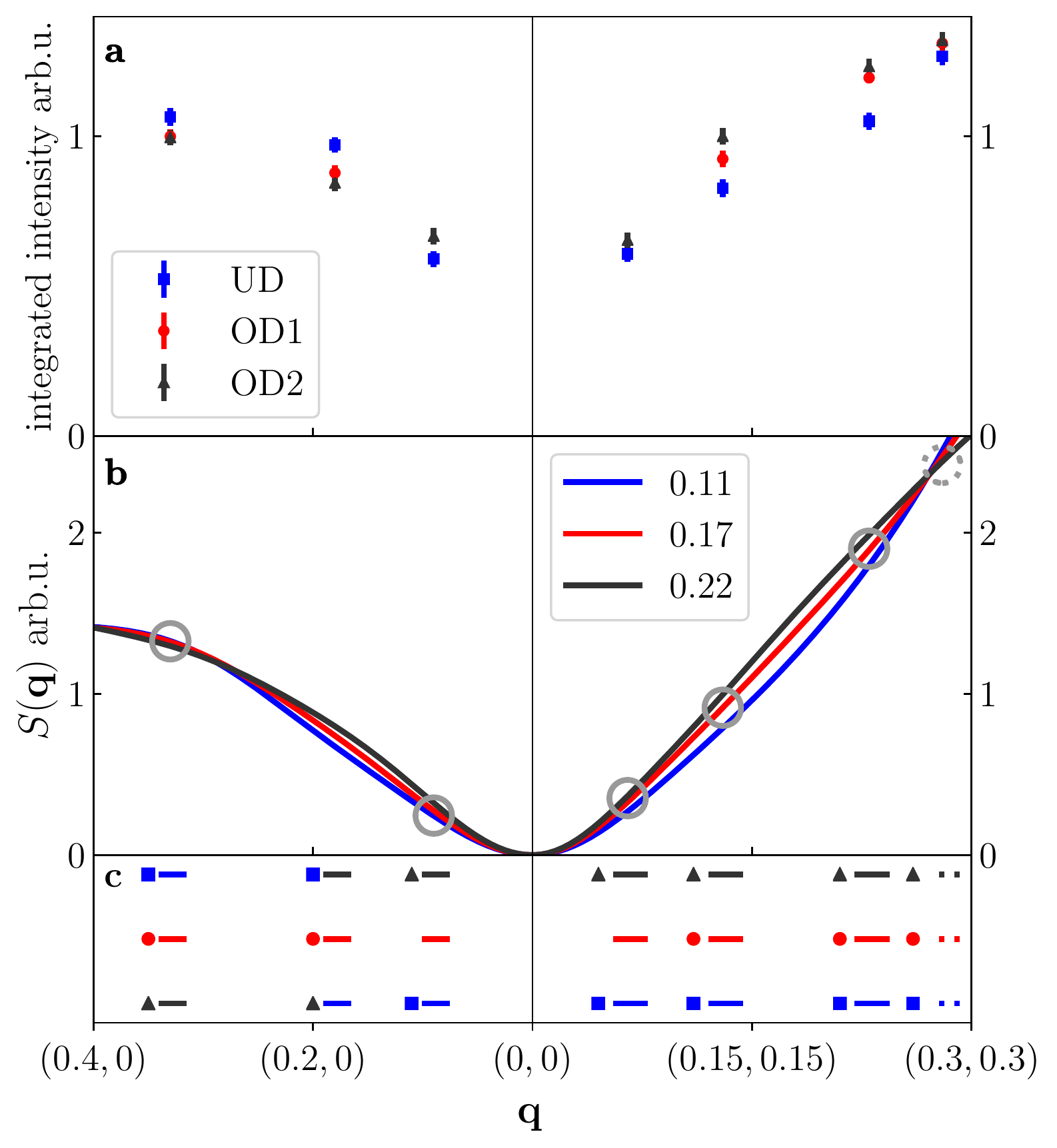}
  \caption{{\bf Comparison between the static spin structure factor in experiment and theory} {\bf a} The experimental proxy for the static spin structure factor, i.e. the integrated intensity of the spin excitations as seen by RIXS on Bi2212, see Fig.~\ref{fig:boat3}{\bf h}. The points are normalised to the value of the intensity at $\mathbf{q}=(0.13,0.13)$ for the OD2 doping level. The error bars represent the integration errors assuming that the spectral errors have an energy correlation defined by the resolution function (see Methods). {\bf b} Static spin structure factor $S(\mathbf{q})$ obtained using DMRG on a $6\times 6$ cluster (see text for further details) for the $t$-$t'$-$t''$-$J$ model and three different hole-doping levels. The results are normalised to the value of $S(\mathbf{q})$ at $\mathbf{q}\simeq(0.13,0.13)$ for the doping level $n=0.22$. Model parameters: $J=0.4t$, $t'=-0.3t$, $t''=0.15t$. {\bf c} Schematic comparison between the face-value of the experimental and numerical sequences of intensities: points refer to RIXS experiments and lines to the DMRG numerical results. The sequence at $\mathbf q = (0.28, 0.28)$ momentum refer to the sequence seen in the numerical results at $\mathbf q = (0.267, 0.267)$, see text for further details. Only $\mathbf q=(0.18,0)$ momentum cannot be reproduced by our theoretical calculations.
  }
  \label{fig:boat5}
\end{figure}

We now discuss how our theoretical calculations compare to the experimental results presented in Fig.~\ref{fig:boat3}. Our aim within the calculations is to reproduce the switching in the sequence of intensities upon doping. Hence, our focus is on this qualitative aspect of the experimental results, rather than on the quantitative reproduction of the experimental data within our theoretical calculations. The main results are shown in Fig.~\ref{fig:boat5}: whereas Fig.~\ref{fig:boat5}a presents the experimental integrated intensity of spin spectral weights for the three measured doping levels UD, OD1, and OD2 (see above), Fig.~\ref{fig:boat5}b shows the calculated $S(\mathbf{q})$ with the $t$-$t'$-$t''$-$J$  model in the restricted Brillouin zone kinematically available to the experiments. Finally, we schematically compare the doping evolution of the experimental and theoretical intensities at all experimental momenta in Fig.~\ref{fig:boat5}c. The crucial message here is that, overall, there is good \textit{qualitative} agreement between theory and experiment. First, the experimentally observed small anisotropy between the ($\uppi$, $\uppi$) and ($\uppi$, 0) directions---the  ($\uppi$, $\uppi$) direction shows larger intensities than ($\uppi$, 0) at high $\mathbf q$---is also reproduced by our calculations, although it is not as small as in the RIXS experiment. Second, at five crucial momentum points the theoretical calculations give the same sequence of intensities of the spin structure factor as a function of doping as the face value of the experimental results (see five grey circles in Fig.~\ref{fig:boat5}b and Fig.~\ref{fig:boat5}c). In addition, the sequence of intensities at $\mathbf q = (0.28,0.28)$ can be reproduced with a slightly smaller momentum (see dashed grey circle in Fig.~\ref{fig:boat5}b and dashed lines in Fig.~\ref{fig:boat5}c). This indicates that a small fine-tuning of the model parameters might give a full agreement between theory and experiment also at this momentum. Last but not least we note that the aforementioned sequence of intensities, and therefore the agreement with the theoretical results, is largely confirmed also when the experimental error bars are taken into account (see Supplementary Note 2).

Besides the overall agreement, there are two important discrepancies between the experiments and calculations. First, the experimentally observed sequence of intensities at $\mathbf q = (0.18, 0)$ is not reproduced by the calculations. Although the calculations do produce a crossing to a decreasing sequence upon doping at a relatively large momentum [${\bf q}>(0.28, 0)$], which satisfies the experimental results at $\mathbf q = (0.33, 0)$, they fail to reproduce the intensity sequence at the intermediate momentum $\mathbf q = (0.18, 0)$. The experimental results thus indicate this crossing should happen at a smaller momentum [${\bf q}< (0.18, 0)$]. Within the $t$-$J$-like model and the parameter sets we considered in this work, the current choice with both $t'$ and $t''$ gives the best agreement (see discussion on Fig.~\ref{fig:boat6} below). To fully reconcile this discrepancy, one may need to further fine-tune the Hamiltonian parameters, and in particular, consider their doping-dependence \cite{Drozdov2018}.

Second, the slope of the momentum dependence of the integrated experimental intensities is smaller compared with the calculated $S(\mathbf{q})$. The slope extrapolates to a non-zero residual intensity at the Brillouin zone center ${\bf q} = (0, 0)$, making it gap-like, which is not found in the numerical results. The reason for this apparent disagreement is likely the inter-layer interaction in this bilayer cuprate, which gives rise to the onset of the optical and acoustic magnon branches \cite{Reznik1996, Hayden1996, LeTacon2011, Peng2017}. While the 2D antiferromagnetic acoustic spin wave has a zero structure factor at the zone center [${\bf q} = (0, 0)$], the optical branch due to the inter-layer coupling will show a non-zero intensity, thus leading to a `gap' in the zone center. As our calculations do not include the coupling between the layers, this gap feature is absent. However, due to the inter-layer interaction being much smaller than the intra-layer one, the two branches quickly merge as $\mathbf q$ moves away from the zone center ($\gtrsim$ 0.1 r.l.u.) \cite{Reznik1996, Hayden1996, LeTacon2011}, and their total intensities will be dominated by the intra-layer parameters. Therefore, the comparison between our calculations and the experiments is in reasonable agreement apart from the zone center.

\begin{figure}[t]
\centering
  \includegraphics[width=\linewidth]{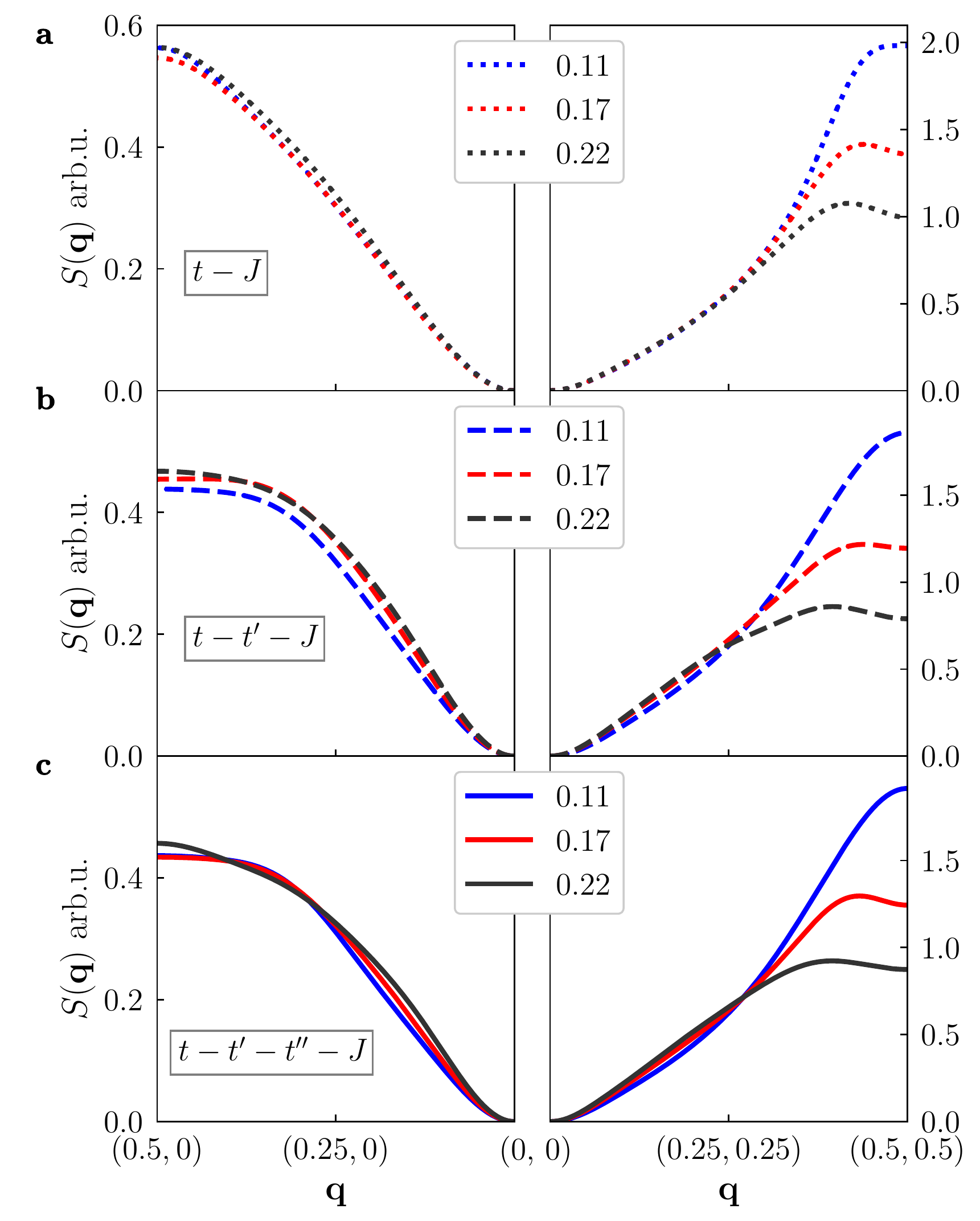}
  \caption{{\bf Theoretical static spin structure factor and the longer-range electronic hoppings.} Static spin structure factor $S(\mathbf{q})$ obtained using DMRG on a $6\times 6$ cluster (see text for further details) for three different hole-doping levels and {\bf a} the $t$-$J$ model, {\bf b} the $t$-$t'$-$J$, and {\bf c} the $t$-$t'$-$t''$-$J$ model. Model parameters as in Fig.~\ref{fig:boat5}. Note the enlarged momentum coverage w.r.t. Fig.~\ref{fig:boat5} and different scales of $S(\mathbf q)$ for the ($\uppi$, $\uppi$) and ($\uppi$, 0) directions of the Brillouin zone.}
  \label{fig:boat6}
\end{figure}

The agreement between the $t$-$t'$-$t''$-$J$ model and experiments can be appreciated even more after looking at Fig.~\ref{fig:boat6}, where we present the comparison between the theoretical results of the three different $t$-$J$-like models. Fig.~\ref{fig:boat6}a shows $S(\mathbf{q})$ calculated for the `bare' $t$-$J$ model. Within this model, the sequence of intensities of the spin structure factor as a function of doping does not change in the experimental momentum range, unlike the results of the experiments. Fig.~\ref{fig:boat6}b shows the same quantity as Fig.~\ref{fig:boat6}a, but for the $t$-$t'$-$J$ model. In this case, the calculated $S(\mathbf{q})$ in the ($\uppi$, $\uppi$) direction shows the same qualitative behavior as in the experimental case. However, the spin structure factor $S(\mathbf{q})$ in ($\uppi$, 0) direction keeps increasing with doping in the studied momentum range, which is still inconsistent with the experiments. A different parameter choice with a larger value of $t'$ in the $t$-$t'$-$J$ model also does not improve the agreement between calculations and experiments (See Supplementary Figure 4 and Note 4). The decreasing sequence of intensity upon doping in the ($\uppi$, 0) direction is only achieved after including the third neighbor hopping $t''$ as shown in Fig.~\ref{fig:boat6}c. However, it appears at relatively larger $\mathbf{q}$ than the experimental results. Nevertheless, it suggests the importance of long range hoppings in achieving better agreement between experiment and theory. Further improvements may require fine-tuning of the parameters (see above).

\begin{figure}[t]
\centering
  \includegraphics[width=0.9\linewidth]{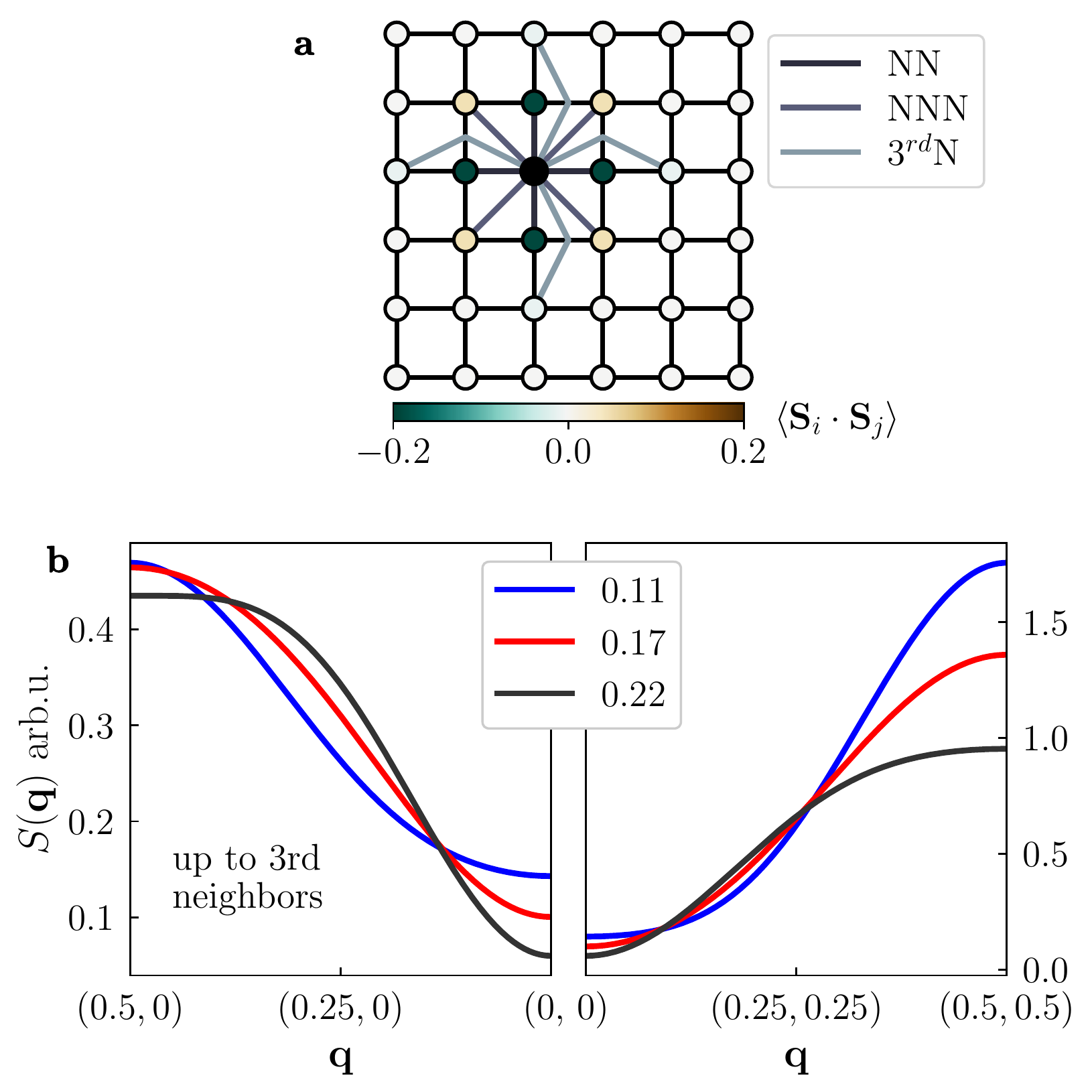}
  \caption{{\bf Theoretical static spin structure factor and the short-range magnetic correlations.} {\bf a} The $6\times 6$ cluster used in the DMRG calculations and the nearest neighbor (NN), next nearest neighbor (NNN) and third neighbor ($3^{rd}$N) spin bonds; the color scale applied to the sites shows the value of the average spin-spin correlation for those particular neighbors. {\bf b} Static spin structure factor $S(\mathbf{q})$ obtained using DMRG on a $6\times 6$ cluster (see text for further details) for three different hole-doping levels and for the $t$-$t'$-$t''$-$J$ model (parameters as in Fig.~\ref{fig:boat5}) with keeping only first, second and third neighbor spin-spin correlations in the Fourier Transform of Eq. (5). Note the enlarged momentum coverage w.r.t. Fig.~\ref{fig:boat5} and different scales of  $S(\mathbf{q})$ for the ($\uppi$, $\uppi$) and ($\uppi$, 0) directions of the Brillouin zone.}
  \label{fig:boat7}
\end{figure}

Due to our real space approach, we have full control over the contribution of spin correlations of \textit{different range} to the static spin structure factor. In fact, we can cut the summation in the Fourier transform [cf.~Eqs.~(5-6) in the Methods] to include only up to a certain number of neighbors. This analysis allows us to investigate the least effective range of magnetic correlations that can qualitatively produce the spin structure factor $S(\mathbf{q})$ upon doping. In Fig.~\ref{fig:boat7}, we show the main results of this analysis on the $t$-$t'$-$t''$-$J$  model. Fig.~\ref{fig:boat7}a is a cartoon description of the real space spin-spin correlations between one sample site around the centre of the $6\times6$ cluster and its neighbors up to third-nearest ones. The color scale represents the value of such real space correlations. In Fig.~\ref{fig:boat7}b, we plot the spin structure factor $S(\mathbf{q})$ calculated by including only up to third-neighbor spin-spin correlations. Since the short-range correlations mostly account for the large-$\mathbf{q}$ dynamics, they become poor in sketching the small-$\mathbf{q}$ properties, which leaves an artificial gap around $\mathbf{q} = (0, 0)$. Except the small-$\mathbf{q}$ region, where the gap appears, the results show an ascending intensity as a function of doping in the intermediate $\mathbf{q}$ range in both the ($\uppi$, $\uppi$) and ($\uppi$, 0) directions and a crossover to descending sequence in larger $\mathbf{q}$, which qualitatively agree with the spin structure factor $S(\mathbf{q})$ calculated from the full Fourier transform depicted in Fig.~\ref{fig:boat6}c. We also notice that almost all of the spectral weight of $S(\mathbf{q})$ at ($\uppi$, $\uppi$) is already accounted for once solely the short-range magnetic correlations up to the third neighbors are taken into account, cf. Fig.~\ref{fig:boat5}b and Fig.~\ref{fig:boat7}c. This is in stark contrast with the undoped case (see Supplementary Figure 3): in the latter case, the spectral weight at ($\uppi$, $\uppi$) is strongly underestimated when only the short-range magnetic correlations are taken into account. This is due to the importance of long-range correlations in the ordered antiferromagnet stabilized at half-filling. An in-depth discussion of these properties, as well as a similar analysis with different range of correlations for the $t$-$J$ and $t$-$t'$-$J$ models can be found in the Supplementary Note 3.\\

%\section*{DISCUSSION}
\noindent{\bf DISCUSSION}\\
Our main experimental result is the unambiguous assessment of the momentum and doping evolution of the disentangled intrinsic spin and charge excitations measured by Cu $L_3$-edge RIXS in doped Bi2212 samples. The disentangled spin responses show profiles that can be well fitted by a damped harmonic oscillator model, although they become over damped along ($\uppi$, $\uppi$) direction. The momentum and doping dependence of the spin and charge responses are different, implying the distinct nature of the two responses. In addition, the disentangled spin excitations well persist into the over doped sample, which confirms that RIXS indeed observes the persistence of spin excitations in a large part of the Brillouin zone in doped cuprates.

The obtained experimental results on the spin excitations are qualitatively reproduced by numerical simulations of the $t$-$J$-like model. It turns out that the bare $t$-$J$ model is not enough and instead this model has to be supplemented by longer-range hoppings—which points out the decisive role of such hoppings in reproducing the experimentally observed paramagnons in doped cuprates. Furthermore, the extensive real space analysis shows that short-range magnetic correlations are needed in order to cause the observed persistence of spin excitations in doped cuprates, meaning they need to be paramagnonic in nature. From that, two important consequences follow: On one hand, within the class of cuprate models with localized spins, those without any magnetic correlations at all seem not to be realistic for doped cuprates. (We note in passing that the class of models with localized spins is not only restricted to the studied $t$--$J$ models. The Hubbard (or charge transfer) model description of the doped cuprates also possesses localized moments whose value is not substantially reduced compared with the $t$--$J$ model case, since the number of doubly occupied sites is substantially suppressed in the Hubbard-like models with realistic values of the on-site Coulomb repulsion, cf. Fig.~1 of~\cite{Moreo1990} or Fig.~6 of~\cite{Kung2015}.) On the other hand, this means that longer-range magnetic correlations do not play a crucial role in the doped cuprates. Altogether, this helps in resolving the paradox related to the persistence of the spin excitations upon doping the cuprates—despite a rapid collapse of the long-range magnetic correlations.

We also briefly comment on the charge excitations we obtained in the disentanglement analysis. With the same model calculations as for the spin structure factor, we can get the charge structure factor (see Supplementary Figure 5), which can reproduce the increase in intensities as a function of doping in both the ($\uppi$, 0) and ($\uppi$, $\uppi$) directions. However, the intensity anisotropy between the ($\uppi$, 0) and ($\uppi$, $\uppi$) is missing. This might signal the onset of the bimagnons in the charge channel of the RIXS spectrum~\cite{Jia2014, Jia2016} or suggest that the longer-range Coulomb interactions are important~\cite{Fidrysiak2021L}.

On the theory side, there are three important implications of the results presented in this paper: The \textit{first} one is that this work shows that a recent theoretical suggestion that the spin excitations are responsible for the {\it T}-linear dependence of the electronic scattering in the Hubbard model \cite{Wu2021} might indeed become a realistic scenario for the cuprates. This follows from the above-stated conclusion that, without any ambiguities, the collective spin excitations persist in the doped cuprates. The \textit{second} one follows from the suggested crucial role played by the longer-range hoppings in the $t$-$J$ models. Such a result goes in line with, {\it inter alia}, recent works advocating for the strong sensitivity of the phase diagram of the $t$-$J$ like models to the value of the next-nearest neighbor hopping $t'$ \cite{Jiang2019,Jiang2020} and thus `sweetens the bad news' coming from the study suggesting the lack of superconductivity in the ground state of the 2D Hubbard model without longer-range hoppings \cite{Qin2020}.
\textit{The third point} relates to the fact that, as stated above, solely the short-range magnetic correlations are needed to explain the persistence of the intensity of the paramagnons in doped cuprates, similar to the recently established well-defined magnons in the random $t$-$J$ model up to $33\%$ hole doping \cite{Kumar2021}. Thus, an interesting task for theory would be to gain an intuitive understanding of {\it{why}} the short-range magnetic correlation alone can lead to the lack of changes of the paramagnons along the ($\uppi$, 0) direction.

We close the paper by presenting the impact of the results presented above on the issue of pairing mechanism in the cuprate superconductors. While there are many competing theories describing this issue, one of the widely-spread concepts suggests that the pairing is mediated by the magnetic excitations close to ($\uppi$, $\uppi$) momentum~\cite{Scalapino2012}. More precisely, it was shown by {\it inter alia} Nocera {\it et al.}~\cite{Nocera2017} and Huang {\it et al.}~\cite{Huang2017} that the spin fluctuations mediated pairing is mostly due to the low-energy spin excitations with momentum ${\bf q}$ = ($\uppi$, $\uppi$) and that the `doping-persistent' high-energy magnetic excitations away from that momentum, i.e. the ones which are observed by RIXS, are not important to pairing. This finding has been corroborated by the experimental results of Meyers {\it et al.} \cite{Meyers2017} and Dean {\it et al.} \cite{Dean2013}. Interestingly, here we have shown that even the ${\bf q}$ = ($\uppi$, $\uppi$) paramagnons are largely a result of the short-range magnetic correlations. This is because the spectral weight in the magnetic response close to the ${\bf q}$ = ($\uppi$, $\uppi$) momentum is very similar both in the `full' spin structure factor calculations as well as in the ones containing solely the short-range magnetic correlations, see Fig.~\ref{fig:boat6}(c) versus Fig~\ref{fig:boat7}. Note that this is in contrast to the undoped case, for which the peak around ${\bf q}$ = ($\uppi$, $\uppi$) is known to be hugely sensitive to the longer-range magnetic correlations (see Supplementary Figure 3). Since the nature of the ($\uppi$, $\uppi$) paramagnons is central to any of the spin fluctuations mediated superconducting pairing mechanism, this finding plays an important role in our deeper understanding of the puzzle of superconductivity in cuprates. Finally, as a side note on the issue of pairing, the present study stresses the importance of the proper choice of the longer range hoppings ($t'$ and $t''$) in any realistic cuprate modelling. Thus, we believe that any model trying to explain superconductivity in the cuprates should include realistic values of these parameters as slight variations might qualitatively affect the observed physics.
\\

%\section*{METHODS}
\noindent{\bf METHODS}\\
%\subsection*{RIXS Experiments and samples}
\noindent{\bf RIXS Experiments and samples}\\
The RIXS experiments were carried out with the SAXES spectrometer at the ADRESS beamline of the Swiss Light Source at the Paul Scherrer Institut \cite{Ghiringhelli2006,Strocov2010}. The incident X-ray energy was set at the Cu $L_3$ resonance peak at approximately 933 eV. The instrument resolution was determined by the elastic peak measured on carbon tape, giving an overall energy resolution of $\sim$ 100 meV full width half maximum (FWHM). The single-crystal samples of Bi$_2$Sr$_2$CaCu$_2$O$_{8+\delta}$ were prepared by the floating zone method as described in Ref. \cite{Weyeneth2009}. The samples were cleaved at base temperature to present fresh surfaces before the measurements. All the data were collected at base temperature $\sim$ 24 K. The momentum transfer $\mathbf{q}$ is denoted in reciprocal lattice units (r. l. u.) using the pseudo-tetragonal unit cell with a = b = 3.82 {\AA}.\\

\noindent{\bf Azimuthal dependent RIXS measurements}\\
In two-dimensional superconducting cuprates, the single-spin-flip and non-spin-flip excitations in the 3$d_{x^2-y^2}$ orbital as well as the other $d$-$d$ excitations in the Cu $L_3$-edge RIXS spectra show distinct geometry and polarization related characters, which are determined by their different scattering tensors \cite{Ament2011,Ament2009,Haverkort2010}. This allows to assess the nature of these excitations by either resolving the polarizations of both the incident and scattered photons \cite{Fumagalli2019} or by azimuthal dependent RIXS measurements \cite{Kang2019}. In Fig.\ref{fig:boat1}b, we show the scattering geometry and the sample rotation in the azimuthal dependent experiments. The directions of the incident and emitted x-rays are fixed through the scattering angle to 130$^\circ$, thus the total momentum transfer is fixed at $\mathbf{q}$. Two linear polarizations ($\sigma$ and $\uppi$) are used for the incident x-rays while the polarization of the emitted light is not resolved. The plate-like sample is mounted on a wedged sample holder (with wedge angle $\theta_w$ = 10$^\circ$, 20$^\circ$, 40$^\circ$ and 50$^\circ$ in the experiments) to have a certain in-plane momentum transfer. The azimuthal rotation axis is parallel to the total momentum transfer $\mathbf{q}$, so that the projections of $\mathbf{q}$ in the sample reciprocal frame are unchanged during rotation, while the projections of the photon polarization are changing. This allows measuring the azimuthal dependence of the excitations at fixed momentum in the sample momentum space. When rotating the sample, the photon polarization will be continuously rotated in the sample space, and the scattering tensors will then result in different rotation dependences for different excitations.\\

\noindent{\bf Error estimations of the experimental data}\\
The major errors of the RIXS spectra are the statistical errors of the photon counting, which are expressed as the square roots of total photon counts. Another error comes from the uncertainties in determining the zero-energy-loss positions in the spectra, which is done in examining the positions of the elastic peaks. Here we assume that this error is about $\pm5$\% of the resolution, which is about $\pm5$ meV. This error is converted to the error of spectral intensity by multiplying the derivative of the spectrum. Other random errors are accounted for by the fitting errors (95\% confidence interval) in the azimuth-dependent fitting. All the above errors are treated as independent at each energy-loss point and summed to a total error by the square root of their sum of squares. The errors of the integrated intensities are obtained by assuming that the errors of the decomposed spectral functions have an energy correlation defined by the resolution function. The error bars in the fitting of the damped harmonic oscillator (Fig.4) are the fitting errors.\\

\noindent{\bf Fitting by damped harmonic oscillator model}\\
The formula of the damped harmonic oscillator (DHO) model used for the fitting of the spin spectral funtions in Fig.~\ref{fig:boat3} is:
\begin{equation}
    A{\cdot}\frac{\gamma\omega}{(\omega^2-\omega_0^2)^2+4\gamma^2\omega^2}.
\end{equation}
A Gaussian resolution function with 100 meV FWHM is convoluted with the above DHO model to fit the results. The fitting energy range is [-0.8, 0.4] eV for all data except those with the smallest momenta ($\mathbf{q}$ = (0.09, 0) and (0.065, 0.065)), which are fitted in [-0.65, 0.4] eV. This is to reduce the influence of the long tail at high energy loss.\\

\noindent{\bf DMRG calculations}\\
In numerical calculations with OBC cluster, a proper correction is often added into the Hamiltonian to minimize the effects of missing terms at open edge. In this study, we introduced the edge factor to uniformize the mobility of charge.

The correct edge factor is calculated for each doping level and each different model ($t$-$J$, $t$-$t'$-$J$, $t$-$t'$-$t''$-$J$). We extrapolate it by computing the dispersion $\delta=n_{\text{in}}-n_\text{out}$ where $n_\text{in}$ is the averaged electron density taken over the sites which do not belong to the edges and $n_\text{out}$ is the averaged electron density taken over the sites which form the edge of the cluster. We computed the dispersion $\delta$ for different values of the edge factor $\lambda$ for each doping level and model and take the final value of the edge factor $\lambda$ as that at which the dispersion $\delta=0$. Nevertheless, the obtained values ($\lambda \sim 0.9-1.2$) are close enough  to 1 to smoothly connect the inside and edge of the cluster. Furthermore, by introducing this edge factor the Friedel oscillations as well as the most important finite-size effect coming from using OBC can be significantly reduced, as can be seen in the Supplementary Note 6.

We keep up to $m=7000$ states in the DMRG calculations, leading to an error $\epsilon/N=10^{-6}$. We make sure that the local density in the system is isotropic by using the edge factor $\lambda$ as described above and we compute the real space spin-spin correlations $\langle\mathbf S_i \cdot \mathbf S_j\rangle$ for all pairs $(i,j)$ labelling the system's sites. This way, we can compute the spin static factor $S(\mathbf{q})$ as:
\begin{equation}
    S(\mathbf{q})= \frac1N
    \sum_{i,j} \langle \mathbf S_i \cdot \mathbf S_j \rangle \cos(q_x (x_i-x_j)+q_y(y_i-y_j)).
    \label{eq:Sq}
 \end{equation}
This method is only valid when the local $z$-component of the spins is small enough, e.g. $S^z_i \leq 10^{-5}$ for all sites $i$. However, for certain doping levels and models, it was not possible to reach this level of accuracy for the local value of $S^z$. Therefore, we renormalized the $S^zS^z$ correlations and computed $S^z(\mathbf{q})$ as:
\begin{align}
    S^z(\mathbf{q}) &=
    \frac1N \sum_{i,j} (\langle  S^z_i \cdot  S^z_j \rangle-\langle S^z_i\rangle\langle S^z_j\rangle)  \nonumber \\ & \times \cos(q_x (x_i-x_j)+q_y(y_i-y_j)).
\end{align}
Due to the symmetries of the considered models, $S(\mathbf{q})=3S^z(\mathbf{q})$. Equation 6 was used for the doping level $n=0.22$ for both the $t$-$t'$-$J$ and the $t$-$t'$-$t''$-$J$ models (see Fig.~\ref{fig:boat6}).\\

\noindent{\bf DATA AVAILABILITY}\\
 All data needed to evaluate the conclusions in the paper are present in the paper and/or the Supplementary Materials. Additional raw experiment data to reproduce the analysis as well as data and scripts to reproduce the theoretical results are available in~\cite{zenodo}.\\

\noindent{\bf CODE AVAILABILITY}\\
 The code for the numerical calculations will be made available from the corresponding authors upon reasonable request.\\

% References
\medskip
%\bibliographystyle{naturemag}
%\bibliography{bibliography}

% Acknowledgements
\medskip
\noindent \textbf{Acknowledgements}  %delete if not applicable))
The experiments were performed at the ADRESS beamline of the Swiss Light Source at the Paul Scherrer Institut (PSI). The experimental work at PSI is supported by the Swiss National Science Foundation through project no. 200021$\_$178867, and the Sinergia network Mott Physics Beyond the Heisenberg Model (MPBH) (SNSF Research Grants CRSII2$\_$160765/1 and CRSII2$\_$141962). The theoretical work is supported by the Narodowe Centrum Nauki (NCN, Poland) under Project Nos. 2016/22/E/ST3/00560 (C.E.A. and K.W.), 2021/40/C/ST3/00177 (C.E.A.) and 2016/23/B/ST3/00839 (K.W.). We acknowledge support by the Interdisciplinary Centre for Mathematical and Computational Modeling (ICM), University of Warsaw (UW), within grant no G81-4. S.N. acknowledges support from SFB 1143 project A05 (project-id 247310070) of the Deutsche Forschungsgemeinschaft. T.C.A. acknowledges funding from the European Union’s Horizon 2020 research and innovation programme under the Marie Skłodowska-Curie grant agreement No. 701647 (PSI-FELLOW-II-3i program). We acknowledge the valuable discussions with Jonathan Pelliciari, Mingu Kang, Riccardo Comin and Ekaterina P\"arschke. We kindly thank U. Nitzsche for technical support.
For the purpose of Open Access, the author has applied a CC-BY public copyright licence to any
Author Accepted Manuscript (AAM) version arising from this submission\\

\noindent \textbf{Competing Interests:} The authors declare that they have no competing interests.\\

\noindent \textbf{Author Contributions:} W.Z. and C.E.A. contributed equally to this work. W.Z. and T.S. conceived the project. K.W. conceived the theory approach. W.Z., Y.T., T.C.A., E.P. and T.S. performed the RIXS experiments with the support of V.S. C.E.A. performed the DMRG calculations with the support of S.N. E.G. grew and characterized the single crystals. W.Z. and C.E.A. analyzed the data in discussion with T.S., S.N., and K.W. W.Z., C.E.A., K.W., and T.S. wrote the manuscript with the input from all authors. T.S. and K.W. coordinated the research.\\

%\clearpage
%\title{Supplementary Materials for Unravelling the Nature of the Spin Excitations Disentangled From the Charge Contributions in a Doped Cuprate Superconductor}
\renewcommand{\refname}{Supplementary References}
\setcounter{equation}{0}
\renewcommand{\theequation}{S\arabic{equation}}
\renewcommand{\tablename}{Supplementary Table}

\noindent{\bf Supplementary Note 1: Absorption coefficients and self-absorption correction}

Self-absorption is always present in soft X-ray RIXS measurements on bulk samples. It will largely modify the RIXS intensity and has to be considered in azimuthal dependent measurement. To derive its contribution, one needs the knowledge of the linear absorption coefficient $\mu(E, \boldsymbol{\epsilon})$ of the studied sample, which depends on the energy ($E$) of the X-rays as well as the orientation of the polarization vector ($\boldsymbol{\epsilon}$) in the sample space. For a single crystal, the absorption coefficient can be expressed in a tensor form, which is constrained by the point group symmetry of the crystal. The Bi2212 sample has a tetragonal structure, leading to a diagonal absorption tensor with only two independent elements, $f_{aa}(E)$ and $f_{cc}(E)$, which correspond to the absorption coefficients with polarization vector in the sample $a$-$b$ plane and along the out-of-plane $c$ axis, respectively \cite{Bricogne2005}. The absorption coefficient with arbitrary polarization vector is,
\begin{equation}
    \mu(E, \boldsymbol{\epsilon})=\boldsymbol{\epsilon}^T\cdot\begin{pmatrix} f_{aa}(E)&0&0\\0&f_{aa}(E)&0\\0&0&f_{cc}(E) \end{pmatrix}\cdot\boldsymbol{\epsilon}.
\end{equation}

In this study, we use X-ray absorption spectroscopy (XAS) measured by total electron yield (TEY) to evaluate the absorption tensor elements $f_{aa}(E)$ and $f_{cc}(E)$ of the Bi2212 samples around the Cu  $L_3$-edge. The inset of Fig. S1 displays the geometry of the measurements. The in-plane element $f_{aa}(E)$ is probed by $\sigma$ polarized X-rays, while the out-of-plane element $f_{cc}(E)$ is probed at grazing incidence and $\pi$ polarization. As the fully grazing incidence is not possible, here we use 20$^\circ$ incident angle, which gives a result of ${\sin}^2 20^\circ{\cdot}f_{aa}(E)+{\cos}^2 20^\circ{\cdot}f_{cc}(E)$. The TEY signal is usually proportional to the absorption coefficient given that the electron escape depth $L$ is much smaller than the photon penetration depth $\lambda{\cdot}{\sin}\alpha$, where $\lambda$=1/$\mu$ and $\alpha$ is the incident angle \cite{Nakajima1999,Henneken2000,Ruosi2014}. Although $L\ll\lambda$ is usually true in cuprates \cite{Ruosi2014}, ${\sin}\alpha$ will reduce the photon penetration depth when $\alpha$ is small at grazing incidence, which will reduce the proportionality especially at the resonance where $\mu$ is large. Such a saturation effect can be described by the following relation \cite{Nakajima1999,Henneken2000,Ruosi2014}:
\begin{equation}
    \text{TEY}(E, \alpha, \boldsymbol{\epsilon})\propto\frac{ML}{\lambda(E, \boldsymbol{\epsilon})\cdot\sin\alpha}\cdot\frac{1}{1+L/(\lambda(E, \boldsymbol{\epsilon})\cdot\sin\alpha)}.
\end{equation}
where $M$ is a material dependent factor and the second fraction on the right side expresses the saturation effect. One can see that this saturation factor will distort the shape of TEY when (1) $L/[\lambda(E, \boldsymbol{\epsilon})\cdot\sin \alpha]\sim$ 1 and (2) the $\lambda(E, \boldsymbol{\epsilon})$ is strongly energy dependent such as around the absorption edge. To address the possible saturation effect in our samples, we measured the TEY at several different incident angles at $\sigma$ polarization, as shown in Supplementary Figure 1(a) – (c). The saturation effect is relatively small with only a slight intensity reduction on the resonance peak in the UD and OD1 samples but becomes more obvious in the OD2 sample. Supplementary Figure 1(d) shows the incident angle dependence of $\text{TEY}\cdot\sin \alpha$ at the resonance peak and pre-edge background. The solid lines are the fitting by a function of $A/(1+1/(r\cdot\sin \alpha))$, in which $r$ gives an estimate of $\lambda/L$. The fitted values of $r$ are 22, 40, and 11 for the UD, OD1, and OD2 samples, respectively. We note that this is just a rough estimate due to the limited points of measured angles. Nevertheless, these values qualitatively agree with the previous measurement on a YBCO film with $r$$\sim$20 \cite{Ruosi2014}. The results suggest that the TEY at normal incidence is nearly proportional to the absorption coefficient. So we take the TEY at normal incidence and $\sigma$ polarization as the in-plane absorption tensor element $f_{aa}(E)$. For the TEY at $\alpha=20^\circ$ and $\pi$ incidence, some saturation effects exist. However, as the resonance peak amplitude is relatively small ($\sim15\%$) compared to the pre-edge background, the saturation factor is much less energy dependent compared to the $\sigma$ polarization. Thus, we can use an overall energy-independent correction factor determined by the change of the pre-edge background from $\alpha=90^\circ$ to $\alpha=20^\circ$ at $\sigma$ polarization to correct the spectra. By subtracting the contribution of ${\sin}^2 20^\circ{\cdot}f_{aa}(E)$, one can then get the out-of-plane element $f_{cc}(E)$. Supplementary Figure 1(e) shows the obtained $f_{aa}(E)$ and $f_{cc}(E)$ of all three samples, which are normalized to make the pre-edge background the same. As can be seen, there are almost no resonance peaks in $f_{cc}(E)$, which confirms the in-plane 3$d_{x^2-y^2}$ character of the samples. $f_{aa}(E)$ and $f_{cc}(E)$ are almost the same for the three samples, except for a small intensity reduction in the resonance peak of $f_{aa}(E)$ in OD2 sample. As the doped holes mainly go to the oxygen ligands, this main peak should be mostly unaffected while the shoulder at a bit higher energy ($\sim$934.2 eV) increases with the hole doping \cite{Ronay1991,Nucker1995}. The reduction is thus most likely due to some residual saturation effects in normal incidence in the OD2 sample due to its smaller $\lambda/L$ ratio at resonance. In our analysis, we use the same set of $f_{aa}(E)$ and $f_{cc}(E)$ from the OD1 sample for all three samples.

The self-absorption effect in RIXS is described by the following formula \cite{Chabot2010,Achkar2011,Kang2019}:
\begin{equation}
    I_\mathrm{t}^\mathrm{exp}(E, \phi, \boldsymbol{\epsilon}_\mathrm{i})=\sum_{\boldsymbol{\epsilon}_\mathrm{f}}\frac{I_\mathrm{t}(E, \phi, \boldsymbol{\epsilon}_\mathrm{i}, \boldsymbol{\epsilon}_\mathrm{f})}{\mu_\mathrm{i}(E_\mathrm{i}, \phi, \boldsymbol{\epsilon}_\mathrm{i})+\mu_\mathrm{f}(E_\mathrm{f}, \phi, \boldsymbol{\epsilon}_\mathrm{f})\cdot\frac{-\widehat{\mathbf k}_\mathrm{i}\cdot\widehat{\mathbf n}_(\phi)}{\widehat{\mathbf k}_\mathrm{f}\cdot\widehat{\mathbf n}_(\phi)}}.
\end{equation}
Here the $I_\mathrm{t}^\mathrm{exp}(E, \phi, \boldsymbol{\epsilon}_\mathrm{i})$is the experimental ($\mathrm{exp}$) RIXS intensity for a certain type ($\mathrm{t}$) of excitation which can be spin, charge or $d$-$d$ excitations, and $I_\mathrm{t}(E, \phi, \boldsymbol{\epsilon}_\mathrm{i}, \boldsymbol{\epsilon}_\mathrm{f})$ is the intrinsic RIXS intensity with $E = E_\mathrm{f} - E_\mathrm{i}$ the energy loss. The denominator describes the self-absorption effect, where $\mu_\mathrm{i(f)}$ is the absorption coefficient of the incident (emitting) photons, $\widehat{\mathbf k}_\mathrm{i(f)}$ is the unit vector of the photon propagation direction, and $\widehat{\mathbf n}_(\phi)$ is the unit vector normal to the sample surface. As the final polarization ($\boldsymbol{\epsilon}_\mathrm{f}$) is not resolved in this experiment, the measured intensity is a sum of all the possible $\boldsymbol{\epsilon}_\mathrm{f}$ ($\sigma$ and $\pi$ polarizations). The azimuthal $\phi$ dependence of $I_\mathrm{t}(E, \phi, \boldsymbol{\epsilon}_\mathrm{i}, \boldsymbol{\epsilon}_\mathrm{f})$ can be calculated based on the scattering tensor of the studied excitation and a rotation matrix which links the laboratory coordinates and sample coordinates, as already demonstrated in reference \cite{Kang2019}. Fig. 1c –d in the main text show the $\phi$ dependence of $\sum_{\boldsymbol{\epsilon}_\mathrm{f}}I_\mathrm{t}(E, \phi, \boldsymbol{\epsilon}_\mathrm{i}, \boldsymbol{\epsilon}_\mathrm{f})$ for the non-spin-flip and spin-flip excitations. With the knowledge of the absorption tensor of our samples, $\mu_\mathrm{i(f)}(E_\mathrm{i(f)}, \phi, \boldsymbol{\epsilon}_\mathrm{i(f)})$ can be determined precisely, and we can therefore calculate the self-absorption effect and include it in the final azimuthal dependence $I_\mathrm{t}^\mathrm{exp}(E, \phi, \boldsymbol{\epsilon}_\mathrm{i})$ for a certain type of excitation. The results at an energy loss of -0.25 eV are shown in Fig. 1e and f for the non-spin-flip and spin-flip excitations. The final RIXS spectrum is a sum of all possible types of excitations $\sum_\mathrm{t}I_\mathrm{t}^\mathrm{exp}(E, \phi, \boldsymbol{\epsilon}_\mathrm{i})$, and $I_\mathrm{t}^\mathrm{exp}(E, \phi, \boldsymbol{\epsilon}_\mathrm{i})$ can be expressed as a product of the intrinsic spectral weight of the excitation $w_\mathrm{t}(E)$ and the azimuthal dependent geometry factor $A_\mathrm{t}(E, \phi, \boldsymbol{\epsilon}_\mathrm{i})$. In the low-energy part of the spectra in cuprates, with the spin-flip and non-spin-flip excitations related to 3$d_{x^2-y^2}$ orbital dominating, the RIXS intensity is then explained by eq. (1), which can be decomposed into the two components by their azimuthal dependence.\\

\noindent{\bf Supplementary Note 2: Consequences of including the error bars for the doping-dependence of the sequence of experimental spin structure intensities.}

In order to show the consequences of including the error bars for the doping-dependence of the sequence of experimental spin structure intensities, we start by presenting
the error bars of the experimental results for each available momentum and doping level, see Supplementary Table 1.

We note that we define the two spin structure intensities as different if the error bars do not overlap, i.e., their differences are twice the widths of the error bars. Thus, the doping development is clear for five momenta points [$(0.33, 0)$, $(0,18, 0)$, $(0.09, 0)$, $(0.13, 0.13)$ and $(0.23, 0.23)$] when the error bars are taken into account. Although at some momenta [$(0.33, 0)$ and $(0.18, 0.18)$ $(0.23, 0.23)$], the error bars of OD1 and OD2 are overlapped, they are quite well separated with UD ones, which means there is still clear doping changes from UD to OD2. The intensities of different doping at (0.065, 0.065) and (0.28, 0.28) are quite similar within the error bars, but they are consistent with the small difference in the calculated $S({\bf q})$.\\

\begin{table*}[t]
\caption{Error bars of  experimental spin structure factors (integrated spin excitation intensities obtained from RIXS) given in the same units as the integrated spin excitation intensities of Fig.~5a of the main text. \label{table:I}}
\begin{tabular}{ p{1.4cm}|p{1.6cm}|p{1.6cm}|p{1.6cm}|p{1.6cm}|p{1.6cm}|p{1.6cm}|p{1.6cm}| }
 \cline{2-8}
 &\multicolumn{7}{|c|}{Momentum} \\
 \hline
 \multicolumn{1}{|c|}{Doping} & (0.33, 0) & (0.18, 0) & (0.09, 0) & (0.065, 0.065) & (0.13, 0.13) & (0.23, 0.23) & (0.28, 0.28)\\
 \hline
 \multicolumn{1}{|c|}{UD} & 0.0305 & 0.0254 & 0.0259 & 0.0273 & 0.0302 & 0.0287 & 0.0302\\
 \hline
 \multicolumn{1}{|c|}{OD1} & 0.0247 & 0.0264 &  &  & 0.0269 & 0.0190 & 0.0294\\
 \hline
 \multicolumn{1}{|c|}{OD2} & 0.0254 & 0.0280 & 0.0267 & 0.0304 & 0.0271 & 0.0256 & 0.0295\\
 \hline
 \end{tabular}
 \end{table*}

\noindent{\bf Supplementary Note 3: Detailed discussions of the influence of the short-range magnetic correlations and longer-range electronic hoppings on the static spin structure factor}

In what follows we study the impact of including solely a restricted number of short-range spin-spin correlations in the Fourier transform defining the static spin structure factor $S ({\bf q})$, see Eq. (5) in the main text of the paper. The main results are presented in Supplementary Figure 2 which shows $S ({\bf q})$ calculated in the three distinct $t$-$J$--like models considered in this paper {\it and} considering up to the third neighbor spin-spin correlations in the Fourier transform. (In the main text, we discuss the case of the $t\text{-}t'\text{-}t''\text{-}J$ model considering up to the third neighbor spin-spin correlations.)

First of all, let us look at the results shown in the first column of Supplementary Figure 2, i.e. once solely spin-spin correlations up to first-neighbors are included (by definition this includes also on site spin-spin correlations) when calculating $S ({\bf q})$. Except for small quantitative differences, the three considered versions of the $t$-$J$ model show the same behavior, meaning the effect of the longer range hopping $t'$ and $t''$ is minimal when the cut-off in the Fourier transform is on the nearest-neighbor correlations. This can be understood when one realizes that first neighbors correlations are always relatively large and antiferromagnetic and the subtle effects induced by the longer range hopping terms are minimal. Moreover, due to the fact that the considered correlations are antiferromagnetic (i.e. negative), a negative spectral weight sets in for small ${\bf q}$ in $S(\mathbf q)$. We would like to underline that this result is not linked to the instability of the ground state, as also the ground state in these calculations is still the exact ground state of each considered model; instead, this is due to the approximate calculations of $S ({\bf q})$, namely the respective cuts in the Fourier series (see above). As a side remark let us note that when one considers up to first neighbors in $S(\mathbf q)$, its behavior is qualitatively the same in the nodal and anti-nodal directions of the Brillouin zone. In particular, the damping of the peak at $(\pi, \pi)$ is recovered, but the same behavior happens at $(\pi, 0)$.

Next, we move on to the second column of Supplementary Figure 2, where we consider up to second-neighbor spin-spin correlations in the Fourier transform for $S ({\bf q})$. Now the two directions in the Brilloin zone, $(\pi, \pi)$ and $(\pi, 0)$, show different behaviors. The damping of the $(\pi, \pi)$ peak is more prominent and the intensity at this point is increased for all three models. No negative weight is present. The main features present in the $S(\mathbf q)$ are already present for the $t \text - J$ model. A distinct behavior is now visible when longer range hoppings are included: in the $(\pi, 0)$ direction the different doping lines do not cross nor overlap, while they do so once the ``bare'' $t \text - J$ model is considered. In the $(\pi, \pi)$ direction, both the $t \text - t' \text - J$ and $t \text - t' \text - t'' \text - J$ models already show the onset of a line crossing around $(\pi/2, \pi/2)$. However, as previously stated, no crossing is present in the $(\pi, 0)$ direction, meaning we need to include further neighbor spin correlations in the Fourier transform to recover this behavior.

Finally, we focus on the third column of Supplementary Figure 2, where up to third-neighbor spin-spin correlations are included in the Fourier transform. When considering the simpler $t \text - J$ model, there is again an onset of negative weight, which has disappeared in the previous case. This comes from the inclusion of third-neighbor spin-spin correlations, which are antiferromagnetic, i.e., negative. Being quite strong, they are not compensated by the ferromagnetic, i.e., positive, second-neighbors correlations, which leads effectively to a negative weight. On the other hand, the intensity of the $(\pi, \pi)$ peak is now doubled compared to the one seen when only first-neighbor correlations are included. Qualitatively, there is not a big difference if compared to the previous case where up to second-neighbors correlations are being considered. If longer range hoppings are included, the negative weight problem is solved. Let us now focus on the $t \text - t' \text - J$ case: in the $(\pi, \pi)$ direction, the intensity of the peak has increased and it is comparable with that of the full $S(\mathbf q)$. Furthermore, the crossing at $(\pi/2, \pi/2)$ is present. If we now look at the $(\pi,0)$ direction, we see the onset of a crossing close to $(\pi,0)$. Compared to the full $S(\mathbf q)$ case [see Fig. 6b], the 0.11 doping level is now ``meeting'' the other doping level lines at $(\pi, 0)$. The last case is the one which includes also the $t''$ hopping as already shown in the main text. The behavior in the $(\pi,\pi)$ direction is similar to that seen in the $t \text - t' \text - J$ model just discussed, therefore we will only examine what happens in the $(\pi, 0)$ direction. Here the crossing close to $(\pi, 0)$ is clearly visible and includes all three doping levels, as seen in experiments and in the full $S(\mathbf q)$.

Lastly, we would like to comment on the presence of a gap at $(0,0)$ momentum transfer observed in the results for both the second and third column of Supplementary Figure 2: it is clear that when one considers the long-range distances in real spaces, this translates to small $\mathbf q$ values in $q$-space. Therefore, the ``wrong'' behavior seen at small $\mathbf q$ is expected to improve more and more when further neighbors are included.

In conclusion, in order to understand the behavior of the integrated intensity $S(\mathbf q)$, it is important to include longer range hopping up to third neighbors, but only shorter range spin-spin correlations are needed to recover the main properties seen in the experimental data.

To show that the above analysis is only possible for the doped system, we carried out a similar analysis for the undoped
(half-filled) system. We present results in Supplementary Figure 3. For a better comparison, we set the $S(\mathbf q)$ range to be the same for all panels. This cuts part of the line in Supplementary Figure 3a, where we are showing the results for the Fourier transform carried out on up to the first-neighbor correlation. For small $\mathbf q$, the static spin structure factor turns out to be negative and is thus cut out of the presented figure. Looking at the progression of $S(\mathbf q)$ when further neighbors are included in the Fourier transform, we observe that the full spin structure factor $S(\mathbf q)$ is gradually better reproduced. Nevertheless, even when considering only up to third neighbors in the pure antiferromagnetic case we cannot obtain a good quantitative agreement with the full $S(\mathbf q)$. This is because the spectral weight of the peak at $(\pi,\pi)$ is still only at about half of its value when up to third neighbors are included.\\

\noindent{\bf Supplementary Note 4: Effect of larger $t'$ on the spin static structure factor}

To check the importance of long-range hopping $t''$ in our model, we consider again the $t$-$t'$-$J$ model, but we set $t'=-0.5 t$ rather than $t'=-0.3 t$. While this value is too large for cuprates, it allows us to  understand the trend of the spin static structure factor when $t'$ is changed. We plot our results in Supplementary Figure 4 and show them only in the Brillouin zone available for the experiments. The first thing we notice is that the agreement in the $(\pi,\pi)$ direction is lost. Indeed, the $0.22$ hole-doping line loses intensity at a rather small $\mathbf q$, so that the sequence of intensities at (0.28, 0.28) is completely different between theory and experiments, in contrast to the results shown in Fig.~5{\bf b} of the main text. When looking at the $(\pi,0)$ direction, we see that the 0.22 hole-doping line does cross the other two intensities, becoming the lowest line, which does not happen for $t'=-0.3 t$. However, the disagreement at $\mathbf q=(0.18,0)$ persists also in this model. Hence, when compared to the option of including longer range hopping $t''$, it seems that the latter is to be preferred as the lowering of the intensity of the 0.22 hole-doping case is achieved for larger $\mathbf q$ without compromising the agreement in the $(\pi, \pi)$ direction.\\

\noindent{\bf Supplementary Note 5: Static charge structure factor}

We calculate the static charge structure factor
\begin{equation}
    N(\mathbf{q})= \langle n(-\mathbf{q}) n(\mathbf{q})\rangle=\frac{1}{N}\sum_{i,j} (\langle  n_i n_j \rangle -\langle n_i\rangle\langle n_j\rangle)\mathrm{e}^{i\mathbf{q}(\textbf{r}_i-\textbf{r}_j)}.
\end{equation}
for the $t$-$t'$-$t''$-$J$ model on a $6\times6$ cluster using DMRG. This is done in a similar manner as the static spin structure factor calculated elsewhere in this manuscript; in particular, the applied edge factors for the different doping levels are the same as those applied to compute the static spin structure factor $S(\mathbf q)$.

We plot the charge structure factor using DMRG in the part of the Brillouin zone available to the RIXS experiments and compare these results to the experimentally extrapolated integrated intensities in the non-spin-flip RIXS sector (Supplementary Figure 4). The sequence of intensities at the probed momentum points is qualitatively well reproduced in the DMRG results for $N(\mathbf q)$. We believe that some differences, especially the much larger relative intensity along the antinodal directions observed in the experiment, can {\it either} be a result of other contribution besides the charge ones in the non-spin-flip channel in the experiments (in particular, the bimagnon contributions~\cite{Jia2014, Jia2016})
{\it or} show that the longer-range Coulomb interactions are important here~\cite{Fidrysiak2021L}.\\

\noindent{\bf Supplementary Note 6: Benchmarking the use of the edge factor in the numerical method}

In order to perform calculations on an open cluster, we have introduced the edge factor $\lambda$ as described in the main text. To prove that this method leads to correct results, we calculated the static spin structure factor $ S(\mathbf q)$ on the half-filled $6\times6$ cluster.
The results are shown in Supplementary Figure 3 and fully agree with the textbook behavior of the Heisenberg model on a square lattice at zero temperature [a dominant peak at $(\pi, \pi$)].

Moreover, as mentioned in the main text, using open clusters could provide much better results than periodic clusters if the open edge is managed properly. To confirm this statement, we compare the static spin structure factor $S(q)$ between short and long open chains in the 1D $t$-$J$ chain with $J/t=0.4$ and $n=2/3$, since the result for a large 2D system is not available. In Supplementary Figure 5, the spin static structure factors for $L=6$ and $L=60$ open chains are compared. The result for $L=60$ is expected to be almost identical to taking the thermodynamic limit. For the $L=6$ open cluster, small local potentials $0.15$ and $-0.3$ have been added on the first and second edge sites, respectively, as edge factors used to achieve a uniform density distribution, i.e., $<n_i>=2/3$ for all $i$, $i$ labeling the sites in the chain. We find a good agreement between $S(q)$ with $L=6$ and $L=60$ except for the peak height around $q=2\pi/3$. This discrepancy is caused mainly by an enhanced finite-size effect due to strong quantum fluctuations in 1D. Thus, such discrepancy will be much smaller in 2D systems. The structure factor for an $L=6$ periodic chain is also plotted. It is difficult to compare it with the thermodynamic limit result, because only discrete momenta are allowed. Besides, the artificial enhancement (suppression) of $S(q)$ at $q=2\pi/3$ (at $q\neq2\pi/3$) is clearly seen. Consequently, we can suggest that the use of an (small) open cluster is a reasonable way to capture the overall features of $S(q)$, unless $S(q)$ shows a very complex behavior.\\

\noindent{\bf Supplementary Note 7: Averaged compared to non-averaged correlations}

In the main text as well as in Supplementary Note 3, the results presented are based on the Fourier transform of the averaged values of the different neighbor spin-spin correlations. To support our choice of showing results based on averaged correlations, Supplementary Figure 6 compares the spin static structure factor $S(\mathbf q)$ calculated using the Fourier transform with keeping up to third neighbor spin-spin correlations for the $t$-$t'$-$t''$-$J$ model---with averaged (over the whole cluster) and non-averaged values of the spin-spin correlations. The differences between the two latter cases are almost invisible, suggesting that our choice of considering only averaged correlations does not introduce any additional approximations.

\medskip
%\bibliographystyle{naturemag}
%\bibliography{App-bibliography}

\begin{thebibliography}{10}
\expandafter\ifx\csname url\endcsname\relax
  \def\url#1{\texttt{#1}}\fi
\expandafter\ifx\csname urlprefix\endcsname\relax\def\urlprefix{URL }\fi
\providecommand{\bibinfo}[2]{#2}
\providecommand{\eprint}[2][]{\url{#2}}

\bibitem{Keimer2015}
\bibinfo{author}{Keimer, B.}, \bibinfo{author}{Kivelson, S.~A.},
  \bibinfo{author}{Norman, M.~R.}, \bibinfo{author}{Uchida, S.} \&
  \bibinfo{author}{Zaanen, J.}
\newblock \bibinfo{title}{From quantum matter to high-temperature
  superconductivity in copper oxides}.
\newblock \emph{\bibinfo{journal}{Nature}} \textbf{\bibinfo{volume}{518}},
  \bibinfo{pages}{179--186} (\bibinfo{year}{2015}).

\bibitem{Lee2006}
\bibinfo{author}{Lee, P.~A.}, \bibinfo{author}{Nagaosa, N.} \&
  \bibinfo{author}{Wen, X.-G.}
\newblock \bibinfo{title}{{D}oping a {M}ott insulator: {P}hysics of
  high-temperature superconductivity}.
\newblock \emph{\bibinfo{journal}{Rev. Mod. Phys.}}
  \textbf{\bibinfo{volume}{78}}, \bibinfo{pages}{17--85}
  (\bibinfo{year}{2006}).

\bibitem{Xu2009}
\bibinfo{author}{Xu, G.} \emph{et~al.}
\newblock \bibinfo{title}{Testing the itinerancy of spin dynamics in
  superconducting {Bi}$_2${Sr}$_2${Ca}{Cu}$_2${O}$_{8+\delta}$}.
\newblock \emph{\bibinfo{journal}{Nat. Phys.}} \textbf{\bibinfo{volume}{5}},
  \bibinfo{pages}{642--646} (\bibinfo{year}{2009}).

\bibitem{Damascelli2003}
\bibinfo{author}{Damascelli, A.}, \bibinfo{author}{Hussain, Z.} \&
  \bibinfo{author}{Shen, Z.-X.}
\newblock \bibinfo{title}{Angle-resolved photoemission studies of the cuprate
  superconductors}.
\newblock \emph{\bibinfo{journal}{Rev. Mod. Phys.}}
  \textbf{\bibinfo{volume}{75}}, \bibinfo{pages}{473--541}
  (\bibinfo{year}{2003}).

\bibitem{Tranquada2013}
\bibinfo{author}{Tranquada, J.~M.}
\newblock \bibinfo{title}{Spins, stripes, and superconductivity in hole-doped
  cuprates}.
\newblock \emph{\bibinfo{journal}{AIP Conference Proceedings}}
  \textbf{\bibinfo{volume}{1550}}, \bibinfo{pages}{114--187}
  (\bibinfo{year}{2013}).

\bibitem{Braicovich2010}
\bibinfo{author}{Braicovich, L.} \emph{et~al.}
\newblock \bibinfo{title}{Magnetic excitations and phase separation in the
  underdoped
  {${\mathrm{La}}_{2\ensuremath{-}x}{\mathrm{Sr}}_{x}{\mathrm{CuO}}_{4}$}
  superconductor measured by resonant inelastic {X}-{R}ay scattering}.
\newblock \emph{\bibinfo{journal}{Phys. Rev. Lett.}}
  \textbf{\bibinfo{volume}{104}}, \bibinfo{pages}{077002}
  (\bibinfo{year}{2010}).

\bibitem{LeTacon2011}
\bibinfo{author}{Le~Tacon, M.} \emph{et~al.}
\newblock \bibinfo{title}{Intense paramagnon excitations in a large family of
  high-temperature superconductors}.
\newblock \emph{\bibinfo{journal}{Nat. Phys.}} \textbf{\bibinfo{volume}{7}},
  \bibinfo{pages}{725--730} (\bibinfo{year}{2011}).

\bibitem{Dean2013}
\bibinfo{author}{Dean, M. P.~M.} \emph{et~al.}
\newblock \bibinfo{title}{Persistence of magnetic excitations in
  {La}$_{2-x}${Sr}$_x${CuO}$_4$ from the undoped insulator to the heavily
  overdoped non-superconducting metal}.
\newblock \emph{\bibinfo{journal}{Nat. Mater.}} \textbf{\bibinfo{volume}{12}},
  \bibinfo{pages}{1019--1023} (\bibinfo{year}{2013}).

\bibitem{Dean2013PRL}
\bibinfo{author}{Dean, M. P.~M.} \emph{et~al.}
\newblock \bibinfo{title}{High-energy magnetic excitations in the cuprate
  superconductor
  {${\mathrm{Bi}}_{2}{\mathrm{Sr}}_{2}{\mathrm{CaCu}}_{2}{\mathrm{O}}_{8\mathrm{+}\ensuremath{\delta}}$}:
  Towards a unified description of its electronic and magnetic degrees of
  freedom}.
\newblock \emph{\bibinfo{journal}{Phys. Rev. Lett.}}
  \textbf{\bibinfo{volume}{110}}, \bibinfo{pages}{147001}
  (\bibinfo{year}{2013}).

\bibitem{Dean2013PRB}
\bibinfo{author}{Dean, M. P.~M.} \emph{et~al.}
\newblock \bibinfo{title}{Magnetic excitations in stripe-ordered
  {L}a${}_{1.875}${B}a${}_{0.125}${C}u{O}${}_{4}$ studied using resonant
  inelastic x-ray scattering}.
\newblock \emph{\bibinfo{journal}{Phys. Rev. B}} \textbf{\bibinfo{volume}{88}},
  \bibinfo{pages}{020403(R)} (\bibinfo{year}{2013}).

\bibitem{LeTacon2013}
\bibinfo{author}{Le~Tacon, M.} \emph{et~al.}
\newblock \bibinfo{title}{Dispersive spin excitations in highly overdoped
  cuprates revealed by resonant inelastic {X}-{R}ay scattering}.
\newblock \emph{\bibinfo{journal}{Phys. Rev. B}} \textbf{\bibinfo{volume}{88}},
  \bibinfo{pages}{020501(R)} (\bibinfo{year}{2013}).

\bibitem{Ishii2014}
\bibinfo{author}{Ishii, K.} \emph{et~al.}
\newblock \bibinfo{title}{High-energy spin and charge excitations in
  electron-doped copper oxide superconductors}.
\newblock \emph{\bibinfo{journal}{Nat. Commun.}} \textbf{\bibinfo{volume}{5}},
  \bibinfo{pages}{3714} (\bibinfo{year}{2014}).

\bibitem{Lee2014}
\bibinfo{author}{Lee, W.~S.} \emph{et~al.}
\newblock \bibinfo{title}{Asymmetry of collective excitations in electron- and
  hole-doped cuprate superconductors}.
\newblock \emph{\bibinfo{journal}{Nat. Phys.}} \textbf{\bibinfo{volume}{10}},
  \bibinfo{pages}{883--889} (\bibinfo{year}{2014}).

\bibitem{Dean2014}
\bibinfo{author}{Dean, M. P.~M.} \emph{et~al.}
\newblock \bibinfo{title}{Itinerant effects and enhanced magnetic interactions
  in {B}i-based multilayer cuprates}.
\newblock \emph{\bibinfo{journal}{Phys. Rev. B}} \textbf{\bibinfo{volume}{90}},
  \bibinfo{pages}{220506(R)} (\bibinfo{year}{2014}).

\bibitem{Guarise2014}
\bibinfo{author}{Guarise, M.} \emph{et~al.}
\newblock \bibinfo{title}{Anisotropic softening of magnetic excitations along
  the nodal direction in superconducting cuprates}.
\newblock \emph{\bibinfo{journal}{Nat. Commun.}} \textbf{\bibinfo{volume}{5}},
  \bibinfo{pages}{5760} (\bibinfo{year}{2014}).

\bibitem{Minola2015}
\bibinfo{author}{Minola, M.} \emph{et~al.}
\newblock \bibinfo{title}{Collective nature of spin excitations in
  superconducting cuprates probed by resonant inelastic {X}-{R}ay scattering}.
\newblock \emph{\bibinfo{journal}{Phys. Rev. Lett.}}
  \textbf{\bibinfo{volume}{114}}, \bibinfo{pages}{217003}
  (\bibinfo{year}{2015}).

\bibitem{Wakimoto2015}
\bibinfo{author}{Wakimoto, S.} \emph{et~al.}
\newblock \bibinfo{title}{High-energy magnetic excitations in overdoped
  {${\mathrm{La}}_{2\ensuremath{-}x}{\mathrm{Sr}}_{x}{\mathrm{CuO}}_{4}$}
  studied by neutron and resonant inelastic {X}-{R}ay scattering}.
\newblock \emph{\bibinfo{journal}{Phys. Rev. B}} \textbf{\bibinfo{volume}{91}},
  \bibinfo{pages}{184513} (\bibinfo{year}{2015}).

\bibitem{Peng2015}
\bibinfo{author}{Peng, Y.~Y.} \emph{et~al.}
\newblock \bibinfo{title}{Magnetic excitations and phonons simultaneously
  studied by resonant inelastic {X}-{R}ay scattering in optimally doped
  {${\mathrm{Bi}}_{1.5}{\mathrm{Pb}}_{0.55}{\mathrm{Sr}}_{1.6}{\mathrm{La}}_{0.4}{\mathrm{CuO}}_{6+\ensuremath{\delta}}$}}.
\newblock \emph{\bibinfo{journal}{Phys. Rev. B}} \textbf{\bibinfo{volume}{92}},
  \bibinfo{pages}{064517} (\bibinfo{year}{2015}).

\bibitem{Ellis2015}
\bibinfo{author}{Ellis, D.~S.} \emph{et~al.}
\newblock \bibinfo{title}{Correlation of the superconducting critical
  temperature with spin and orbital excitations in
  {$({\mathrm{Ca}}_{x}{\mathrm{La}}_{1\ensuremath{-}x})({\mathrm{Ba}}_{1.75\ensuremath{-}x}\text{La}_{0.25+x}){\mathrm{Cu}}_{3}{\mathrm{O}}_{y}$}
  as measured by resonant inelastic {X}-{R}ay scattering}.
\newblock \emph{\bibinfo{journal}{Phys. Rev. B}} \textbf{\bibinfo{volume}{92}},
  \bibinfo{pages}{104507} (\bibinfo{year}{2015}).

\bibitem{Huang2016}
\bibinfo{author}{Huang, H.~Y.} \emph{et~al.}
\newblock \bibinfo{title}{Raman and fluorescence characteristics of resonant
  inelastic {X}-{R}ay scattering from doped superconducting cuprates}.
\newblock \emph{\bibinfo{journal}{Sci Rep}} \textbf{\bibinfo{volume}{6}},
  \bibinfo{pages}{19657} (\bibinfo{year}{2016}).

\bibitem{Monney2016}
\bibinfo{author}{Monney, C.} \emph{et~al.}
\newblock \bibinfo{title}{Resonant inelastic {X}-{R}ay scattering study of the
  spin and charge excitations in the overdoped superconductor
  {${\mathrm{La}}_{1.77}{\mathrm{Sr}}_{0.23}{\mathrm{CuO}}_{4}$}}.
\newblock \emph{\bibinfo{journal}{Phys. Rev. B}} \textbf{\bibinfo{volume}{93}},
  \bibinfo{pages}{075103} (\bibinfo{year}{2016}).

\bibitem{Meyers2017}
\bibinfo{author}{Meyers, D.} \emph{et~al.}
\newblock \bibinfo{title}{Doping dependence of the magnetic excitations in
  {${\mathrm{La}}_{2\ensuremath{-}x}{\mathrm{Sr}}_{x}{\mathrm{CuO}}_{4}$}}.
\newblock \emph{\bibinfo{journal}{Phys. Rev. B}} \textbf{\bibinfo{volume}{95}},
  \bibinfo{pages}{075139} (\bibinfo{year}{2017}).

\bibitem{Ivashko2017}
\bibinfo{author}{Ivashko, O.} \emph{et~al.}
\newblock \bibinfo{title}{Damped spin excitations in a doped cuprate
  superconductor with orbital hybridization}.
\newblock \emph{\bibinfo{journal}{Phys. Rev. B}} \textbf{\bibinfo{volume}{95}},
  \bibinfo{pages}{214508} (\bibinfo{year}{2017}).

\bibitem{Minola2017}
\bibinfo{author}{Minola, M.} \emph{et~al.}
\newblock \bibinfo{title}{Crossover from collective to incoherent spin
  excitations in superconducting cuprates probed by detuned resonant inelastic
  {X}-{R}ay scattering}.
\newblock \emph{\bibinfo{journal}{Phys. Rev. Lett.}}
  \textbf{\bibinfo{volume}{119}}, \bibinfo{pages}{097001}
  (\bibinfo{year}{2017}).

\bibitem{Chaix2018}
\bibinfo{author}{Chaix, L.} \emph{et~al.}
\newblock \bibinfo{title}{Resonant inelastic {X}-{R}ay scattering studies of
  magnons and bimagnons in the lightly doped cuprate
  {${\mathrm{La}}_{2\ensuremath{-}x}{\mathrm{Sr}}_{x}{\mathrm{CuO}}_{4}$}}.
\newblock \emph{\bibinfo{journal}{Phys. Rev. B}} \textbf{\bibinfo{volume}{97}},
  \bibinfo{pages}{155144} (\bibinfo{year}{2018}).

\bibitem{Peng2018}
\bibinfo{author}{Peng, Y.~Y.} \emph{et~al.}
\newblock \bibinfo{title}{Dispersion, damping, and intensity of spin
  excitations in the monolayer
  {${(\text{Bi,Pb})}_{2}{(\text{Sr,La})}_{2}{\mathrm{CuO}}_{6+\ensuremath{\delta}}$}
  cuprate superconductor family}.
\newblock \emph{\bibinfo{journal}{Phys. Rev. B}} \textbf{\bibinfo{volume}{98}},
  \bibinfo{pages}{144507} (\bibinfo{year}{2018}).

\bibitem{Robarts2019}
\bibinfo{author}{Robarts, H.~C.} \emph{et~al.}
\newblock \bibinfo{title}{Anisotropic damping and wave vector dependent
  susceptibility of the spin fluctuations in
  {${\mathrm{La}}_{2\ensuremath{-}x}{\mathrm{Sr}}_{x}{\mathrm{CuO}}_{4}$}
  studied by resonant inelastic {X}-{R}ay scattering}.
\newblock \emph{\bibinfo{journal}{Phys. Rev. B}}
  \textbf{\bibinfo{volume}{100}}, \bibinfo{pages}{214510}
  (\bibinfo{year}{2019}).

\bibitem{James2012}
\bibinfo{author}{James, A. J.~A.}, \bibinfo{author}{Konik, R.~M.} \&
  \bibinfo{author}{Rice, T.~M.}
\newblock \bibinfo{title}{Magnetic response in the underdoped cuprates}.
\newblock \emph{\bibinfo{journal}{Phys. Rev. B}} \textbf{\bibinfo{volume}{86}},
  \bibinfo{pages}{100508(R)} (\bibinfo{year}{2012}).

\bibitem{Benjamin2014}
\bibinfo{author}{Benjamin, D.}, \bibinfo{author}{Klich, I.} \&
  \bibinfo{author}{Demler, E.}
\newblock \bibinfo{title}{Single-band model of resonant inelastic {X}-{R}ay
  scattering by quasiparticles in high-${T}_{c}$ cuprate superconductors}.
\newblock \emph{\bibinfo{journal}{Phys. Rev. Lett.}}
  \textbf{\bibinfo{volume}{112}}, \bibinfo{pages}{247002}
  (\bibinfo{year}{2014}).

\bibitem{Nagy2016}
\bibinfo{author}{Kan\'asz-Nagy, M.}, \bibinfo{author}{Shi, Y.},
  \bibinfo{author}{Klich, I.} \& \bibinfo{author}{Demler, E.~A.}
\newblock \bibinfo{title}{Resonant inelastic {X}-{R}ay scattering as a probe of
  band structure effects in cuprates}.
\newblock \emph{\bibinfo{journal}{Phys. Rev. B}} \textbf{\bibinfo{volume}{94}},
  \bibinfo{pages}{165127} (\bibinfo{year}{2016}).

\bibitem{Jia2016}
\bibinfo{author}{Jia, C.}, \bibinfo{author}{Wohlfeld, K.},
  \bibinfo{author}{Wang, Y.}, \bibinfo{author}{Moritz, B.} \&
  \bibinfo{author}{Devereaux, T.~P.}
\newblock \bibinfo{title}{Using {RIXS} to uncover elementary charge and spin
  excitations}.
\newblock \emph{\bibinfo{journal}{Phys. Rev. X}} \textbf{\bibinfo{volume}{6}},
  \bibinfo{pages}{021020} (\bibinfo{year}{2016}).

\bibitem{Tsutsui2016}
\bibinfo{author}{Tsutsui, K.} \& \bibinfo{author}{Tohyama, T.}
\newblock \bibinfo{title}{Incident-energy-dependent spectral weight of resonant
  inelastic {X}-{R}ay scattering in doped cuprates}.
\newblock \emph{\bibinfo{journal}{Phys. Rev. B}} \textbf{\bibinfo{volume}{94}},
  \bibinfo{pages}{085144} (\bibinfo{year}{2016}).

\bibitem{Kang2019}
\bibinfo{author}{Kang, M.} \emph{et~al.}
\newblock \bibinfo{title}{Resolving the nature of electronic excitations in
  resonant inelastic {X}-{R}ay scattering}.
\newblock \emph{\bibinfo{journal}{Phys. Rev. B}} \textbf{\bibinfo{volume}{99}},
  \bibinfo{pages}{045105} (\bibinfo{year}{2019}).

\bibitem{Peng2017}
\bibinfo{author}{Peng, Y.~Y.} \emph{et~al.}
\newblock \bibinfo{title}{Influence of apical oxygen on the extent of in-plane
  exchange interaction in cuprate superconductors}.
\newblock \emph{\bibinfo{journal}{Nat. Phys.}} \textbf{\bibinfo{volume}{13}},
  \bibinfo{pages}{1201--1206} (\bibinfo{year}{2017}).

\bibitem{Ament2011}
\bibinfo{author}{Ament, L. J.~P.}, \bibinfo{author}{van Veenendaal, M.},
  \bibinfo{author}{Devereaux, T.~P.}, \bibinfo{author}{Hill, J.~P.} \&
  \bibinfo{author}{van~den Brink, J.}
\newblock \bibinfo{title}{Resonant inelastic {X}-{R}ay scattering studies of
  elementary excitations}.
\newblock \emph{\bibinfo{journal}{Rev. Mod. Phys.}}
  \textbf{\bibinfo{volume}{83}}, \bibinfo{pages}{705--767}
  (\bibinfo{year}{2011}).

\bibitem{Bisogni2012}
\bibinfo{author}{Bisogni, V.} \emph{et~al.}
\newblock \bibinfo{title}{Bimagnon studies in cuprates with resonant inelastic
  {X}-{R}ay scattering at the {O} ${K}$ edge. {I}. {A}ssessment on
  {L}a${}_{2}${C}u{O}${}_{4}$ and comparison with the excitation at {C}u
  ${L}_{3}$ and {C}u ${K}$ edges}.
\newblock \emph{\bibinfo{journal}{Phys. Rev. B}} \textbf{\bibinfo{volume}{85}},
  \bibinfo{pages}{214527} (\bibinfo{year}{2012}).

\bibitem{Devereaux2007}
\bibinfo{author}{Devereaux, T.~P.} \& \bibinfo{author}{Hackl, R.}
\newblock \bibinfo{title}{Inelastic light scattering from correlated
  electrons}.
\newblock \emph{\bibinfo{journal}{Rev. Mod. Phys.}}
  \textbf{\bibinfo{volume}{79}}, \bibinfo{pages}{175--233}
  (\bibinfo{year}{2007}).

\bibitem{Sugai2003}
\bibinfo{author}{Sugai, S.}, \bibinfo{author}{Suzuki, H.},
  \bibinfo{author}{Takayanagi, Y.}, \bibinfo{author}{Hosokawa, T.} \&
  \bibinfo{author}{Hayamizu, N.}
\newblock \bibinfo{title}{Carrier-density-dependent momentum shift of the
  coherent peak and the {LO} phonon mode in p-type high-${T}_{c}$
  superconductors}.
\newblock \emph{\bibinfo{journal}{Phys. Rev. B}} \textbf{\bibinfo{volume}{68}},
  \bibinfo{pages}{184504} (\bibinfo{year}{2003}).

\bibitem{Sugai2013}
\bibinfo{author}{Sugai, S.} \emph{et~al.}
\newblock \bibinfo{title}{Superconducting pairing and the pseudogap in the
  nematic dynamical stripe phase of {L}a$_{2-x}${S}r$_x${CuO}$_4$}.
\newblock \emph{\bibinfo{journal}{J. Phys.: Condens. Matter}}
  \textbf{\bibinfo{volume}{25}}, \bibinfo{pages}{475701}
  (\bibinfo{year}{2013}).

\bibitem{Chao1978}
\bibinfo{author}{Chao, K.~A.}, \bibinfo{author}{Spa{\l}ek, J.} \&
  \bibinfo{author}{Ole{\'{s}}, A.~M.}
\newblock \bibinfo{title}{Canonical perturbation expansion of the {H}ubbard
  model}.
\newblock \emph{\bibinfo{journal}{Phys. Rev. B}} \textbf{\bibinfo{volume}{18}},
  \bibinfo{pages}{3453--3464} (\bibinfo{year}{1978}).

\bibitem{Dellanoy2009}
\bibinfo{author}{Delannoy, J.-Y.~P.}, \bibinfo{author}{Gingras, M. J.~P.},
  \bibinfo{author}{Holdsworth, P. C.~W.} \& \bibinfo{author}{Tremblay,
  A.-M.~S.}
\newblock \bibinfo{title}{Low-energy theory of the
  $t$-$t{\ensuremath{'}}$-$t{\ensuremath{''}}$-${U}$ hubbard model at
  half-filling: Interaction strengths in cuprate superconductors and an
  effective spin-only description of {${\mathrm{La}}_{2}{\mathrm{CuO}}_{4}$}}.
\newblock \emph{\bibinfo{journal}{Phys. Rev. B}} \textbf{\bibinfo{volume}{79}},
  \bibinfo{pages}{235130} (\bibinfo{year}{2009}).

\bibitem{Senechal2004}
\bibinfo{author}{S\'en\'echal, D.} \& \bibinfo{author}{Tremblay, A.-M.~S.}
\newblock \bibinfo{title}{Hot spots and pseudogaps for hole- and electron-doped
  high-temperature superconductors}.
\newblock \emph{\bibinfo{journal}{Phys. Rev. Lett.}}
  \textbf{\bibinfo{volume}{92}}, \bibinfo{pages}{126401}
  (\bibinfo{year}{2004}).

\bibitem{Drozdov2018}
\bibinfo{author}{Drozdov, I.~K.} \emph{et~al.}
\newblock \bibinfo{title}{Phase diagram of {Bi$_2$Sr$_2$CaCu$_2$O$_{8+\delta}$}
  revisited}.
\newblock \emph{\bibinfo{journal}{Nat. Commun.}} \textbf{\bibinfo{volume}{9}},
  \bibinfo{pages}{5210} (\bibinfo{year}{2018}).

\bibitem{Jia2014}
\bibinfo{author}{Jia, C.~J.} \emph{et~al.}
\newblock \bibinfo{title}{Persistent spin excitations in doped antiferromagnets
  revealed by resonant inelastic light scattering}.
\newblock \emph{\bibinfo{journal}{Nat. Commun.}} \textbf{\bibinfo{volume}{5}},
  \bibinfo{pages}{3314} (\bibinfo{year}{2014}).

\bibitem{Eder1995}
\bibinfo{author}{Eder, R.}, \bibinfo{author}{Ohta, Y.} \&
  \bibinfo{author}{Maekawa, S.}
\newblock \bibinfo{title}{Anomalous spin and charge dynamics of the $t$-${J}$
  model at low doping}.
\newblock \emph{\bibinfo{journal}{Phys. Rev. Lett.}}
  \textbf{\bibinfo{volume}{74}}, \bibinfo{pages}{5124--5127}
  (\bibinfo{year}{1995}).

\bibitem{Tohyama1995}
\bibinfo{author}{Tohyama, T.}, \bibinfo{author}{Horsch, P.} \&
  \bibinfo{author}{Maekawa, S.}
\newblock \bibinfo{title}{Spin and charge dynamics of the $t$-${J}$ model}.
\newblock \emph{\bibinfo{journal}{Phys. Rev. Lett.}}
  \textbf{\bibinfo{volume}{74}}, \bibinfo{pages}{980--983}
  (\bibinfo{year}{1995}).

\bibitem{Nocera2017}
\bibinfo{author}{Nocera, A.}, \bibinfo{author}{Patel, N.~D.},
  \bibinfo{author}{Dagotto, E.} \& \bibinfo{author}{Alvarez, G.}
\newblock \bibinfo{title}{Signatures of pairing in the magnetic excitation
  spectrum of strongly correlated two-leg ladders}.
\newblock \emph{\bibinfo{journal}{Phys. Rev. B}} \textbf{\bibinfo{volume}{96}},
  \bibinfo{pages}{205120} (\bibinfo{year}{2017}).

\bibitem{Tohyama2018}
\bibinfo{author}{Tohyama, T.}, \bibinfo{author}{Mori, M.} \&
  \bibinfo{author}{Sota, S.}
\newblock \bibinfo{title}{Dynamical density matrix renormalization group study
  of spin and charge excitations in the four-leg
  $t\text{\ensuremath{-}}{t}^{\ensuremath{'}}\text{\ensuremath{-}}{J}$ ladder}.
\newblock \emph{\bibinfo{journal}{Phys. Rev. B}} \textbf{\bibinfo{volume}{97}},
  \bibinfo{pages}{235137} (\bibinfo{year}{2018}).

\bibitem{Parschke2019}
\bibinfo{author}{P\"arschke, E.~M.} \emph{et~al.}
\newblock \bibinfo{title}{Numerical investigation of spin excitations in a
  doped spin chain}.
\newblock \emph{\bibinfo{journal}{Phys. Rev. B}} \textbf{\bibinfo{volume}{99}},
  \bibinfo{pages}{205102} (\bibinfo{year}{2019}).

\bibitem{Jiang2021}
\bibinfo{author}{Jiang, S.}, \bibinfo{author}{Scalapino, D.~J.} \&
  \bibinfo{author}{White, S.~R.}
\newblock \bibinfo{title}{Ground-state phase diagram of the
  $t-t\ensuremath{'}-{J}$ model}.
\newblock \emph{\bibinfo{journal}{Proc. Natl. Acad. Sci. U.S.A.}}
  \textbf{\bibinfo{volume}{118}}, \bibinfo{pages}{e2109978118}
  (\bibinfo{year}{2021}).

\bibitem{Gong2021}
\bibinfo{author}{Gong, S.}, \bibinfo{author}{Zhu, W.} \&
  \bibinfo{author}{Sheng, D.~N.}
\newblock \bibinfo{title}{Robust $d$-wave superconductivity in the
  square-lattice $t\text{\ensuremath{-}}{J}$ model}.
\newblock \emph{\bibinfo{journal}{Phys. Rev. Lett.}}
  \textbf{\bibinfo{volume}{127}}, \bibinfo{pages}{097003}
  (\bibinfo{year}{2021}).

\bibitem{Jiang2018}
\bibinfo{author}{Jiang, H.-C.}, \bibinfo{author}{Weng, Z.-Y.} \&
  \bibinfo{author}{Kivelson, S.~A.}
\newblock \bibinfo{title}{Superconductivity in the doped
  $\mathit{t}\ensuremath{-}\mathit{J}$ model: Results for four-leg cylinders}.
\newblock \emph{\bibinfo{journal}{Phys. Rev. B}} \textbf{\bibinfo{volume}{98}},
  \bibinfo{pages}{140505} (\bibinfo{year}{2018}).

\bibitem{Reznik1996}
\bibinfo{author}{Reznik, D.} \emph{et~al.}
\newblock \bibinfo{title}{Direct observation of optical magnons in
  {YBa$_2$Cu$_3$O$_{6.2}$}}.
\newblock \emph{\bibinfo{journal}{Phys. Rev. B}} \textbf{\bibinfo{volume}{53}},
  \bibinfo{pages}{R14741(R)} (\bibinfo{year}{1996}).

\bibitem{Hayden1996}
\bibinfo{author}{Hayden, S.~M.}, \bibinfo{author}{Aeppli, G.},
  \bibinfo{author}{Perring, T.~G.}, \bibinfo{author}{Mook, H.~A.} \&
  \bibinfo{author}{Do{\u{g}}an, F.}
\newblock \bibinfo{title}{High-frequency spin waves in
  {YBa$_2$Cu$_3$O$_{6.15}$}}.
\newblock \emph{\bibinfo{journal}{Phys. Rev. B}} \textbf{\bibinfo{volume}{54}},
  \bibinfo{pages}{R6905(R)} (\bibinfo{year}{1996}).

\bibitem{Moreo1990}
\bibinfo{author}{Moreo, A.}, \bibinfo{author}{Scalapino, D.~J.},
  \bibinfo{author}{Sugar, R.~L.}, \bibinfo{author}{White, S.~R.} \&
  \bibinfo{author}{Bickers, N.~E.}
\newblock \bibinfo{title}{Numerical study of the two-dimensional {H}ubbard
  model for various band fillings}.
\newblock \emph{\bibinfo{journal}{Phys. Rev. B}} \textbf{\bibinfo{volume}{41}},
  \bibinfo{pages}{2313--2320} (\bibinfo{year}{1990}).

\bibitem{Kung2015}
\bibinfo{author}{Kung, Y.~F.} \emph{et~al.}
\newblock \bibinfo{title}{Doping evolution of spin and charge excitations in
  the {H}ubbard model}.
\newblock \emph{\bibinfo{journal}{Phys. Rev. B}} \textbf{\bibinfo{volume}{92}},
  \bibinfo{pages}{195108} (\bibinfo{year}{2015}).

\bibitem{Fidrysiak2021L}
\bibinfo{author}{Fidrysiak, M.} \& \bibinfo{author}{Spa\l{}ek, J.}
\newblock \bibinfo{title}{Unified theory of spin and charge excitations in
  high-${T}_\textrm{c}$ cuprate superconductors: A quantitative comparison with
  experiment and interpretation}.
\newblock \emph{\bibinfo{journal}{Phys. Rev. B}}
  \textbf{\bibinfo{volume}{104}}, \bibinfo{pages}{L020510}
  (\bibinfo{year}{2021}).

\bibitem{Wu2021}
\bibinfo{author}{W\'u, W.}, \bibinfo{author}{Wang, X.} \&
  \bibinfo{author}{Tremblay, A.-M.}
\newblock \bibinfo{title}{Non-{F}ermi liquid phase and linear-in-temperature
  scattering rate in overdoped two-dimensional {H}ubbard model}.
\newblock \emph{\bibinfo{journal}{Proc. Natl. Acad. Sci. U.S.A.}}
  \textbf{\bibinfo{volume}{119}}, \bibinfo{pages}{e2115819119}
  (\bibinfo{year}{2022}).

\bibitem{Jiang2019}
\bibinfo{author}{Jiang, H.-C.} \& \bibinfo{author}{Devereaux, T.~P.}
\newblock \bibinfo{title}{Superconductivity in the doped {H}ubbard model and
  its interplay with next-nearest hopping $t'$}.
\newblock \emph{\bibinfo{journal}{Science}} \textbf{\bibinfo{volume}{365}},
  \bibinfo{pages}{1424--1428} (\bibinfo{year}{2019}).

\bibitem{Jiang2020}
\bibinfo{author}{Jiang, Y.-F.}, \bibinfo{author}{Zaanen, J.},
  \bibinfo{author}{Devereaux, T.~P.} \& \bibinfo{author}{Jiang, H.-C.}
\newblock \bibinfo{title}{Ground state phase diagram of the doped hubbard model
  on the four-leg cylinder}.
\newblock \emph{\bibinfo{journal}{Phys. Rev. Research}}
  \textbf{\bibinfo{volume}{2}}, \bibinfo{pages}{033073} (\bibinfo{year}{2020}).

\bibitem{Qin2020}
\bibinfo{author}{Qin, M.} \emph{et~al.}
\newblock \bibinfo{title}{Absence of superconductivity in the pure
  two-dimensional {H}ubbard model}.
\newblock \emph{\bibinfo{journal}{Phys. Rev. X}} \textbf{\bibinfo{volume}{10}},
  \bibinfo{pages}{031016} (\bibinfo{year}{2020}).

\bibitem{Kumar2021}
\bibinfo{author}{Kumar, A.}, \bibinfo{author}{Sachdev, S.} \&
  \bibinfo{author}{Tripathi, V.}
\newblock \bibinfo{title}{Quasiparticle metamorphosis in the random
  $t\text{\ensuremath{-}}{J}$ model}.
\newblock \emph{\bibinfo{journal}{Phys. Rev. B}}
  \textbf{\bibinfo{volume}{106}}, \bibinfo{pages}{L081120}
  (\bibinfo{year}{2022}).

\bibitem{Scalapino2012}
\bibinfo{author}{Scalapino, D.~J.}
\newblock \bibinfo{title}{A common thread: The pairing interaction for
  unconventional superconductors}.
\newblock \emph{\bibinfo{journal}{Rev. Mod. Phys.}}
  \textbf{\bibinfo{volume}{84}}, \bibinfo{pages}{1383--1417}
  (\bibinfo{year}{2012}).

\bibitem{Huang2017}
\bibinfo{author}{Huang, E.~W.}, \bibinfo{author}{Scalapino, D.~J.},
  \bibinfo{author}{Maier, T.~A.}, \bibinfo{author}{Moritz, B.} \&
  \bibinfo{author}{Devereaux, T.~P.}
\newblock \bibinfo{title}{Decrease of $d$-wave pairing strength in spite of the
  persistence of magnetic excitations in the overdoped {H}ubbard model}.
\newblock \emph{\bibinfo{journal}{Phys. Rev. B}} \textbf{\bibinfo{volume}{96}},
  \bibinfo{pages}{020503} (\bibinfo{year}{2017}).

\bibitem{Ghiringhelli2006}
\bibinfo{author}{Ghiringhelli, G.} \emph{et~al.}
\newblock \bibinfo{title}{{SAXES}, a high resolution spectrometer for resonant
  {X}-{R}ay emission in the 400-1600ev energy range}.
\newblock \emph{\bibinfo{journal}{Rev. Sci. Instrum.}}
  \textbf{\bibinfo{volume}{77}}, \bibinfo{pages}{113108}
  (\bibinfo{year}{2006}).

\bibitem{Strocov2010}
\bibinfo{author}{Strocov, V.~N.} \emph{et~al.}
\newblock \bibinfo{title}{{High-resolution soft X-ray beamline ADRESS at the
  Swiss Light Source for resonant inelastic X-ray scattering and angle-resolved
  photoelectron spectroscopies}}.
\newblock \emph{\bibinfo{journal}{J. Synchrotron Rad.}}
  \textbf{\bibinfo{volume}{17}}, \bibinfo{pages}{631--643}
  (\bibinfo{year}{2010}).

\bibitem{Weyeneth2009}
\bibinfo{author}{Weyeneth, S.}, \bibinfo{author}{Schneider, T.} \&
  \bibinfo{author}{Giannini, E.}
\newblock \bibinfo{title}{Evidence for {K}osterlitz-{T}houless and
  three-dimensional ${XY}$ critical behavior in
  {${\text{Bi}}_{2}{\text{Sr}}_{2}{\text{CaCu}}_{2}{\text{O}}_{8+\ensuremath{\delta}}$}}.
\newblock \emph{\bibinfo{journal}{Phys. Rev. B}} \textbf{\bibinfo{volume}{79}},
  \bibinfo{pages}{214504} (\bibinfo{year}{2009}).

\bibitem{Ament2009}
\bibinfo{author}{Ament, L. J.~P.}, \bibinfo{author}{Ghiringhelli, G.},
  \bibinfo{author}{Sala, M.~M.}, \bibinfo{author}{Braicovich, L.} \&
  \bibinfo{author}{van~den Brink, J.}
\newblock \bibinfo{title}{Theoretical demonstration of how the dispersion of
  magnetic excitations in cuprate compounds can be determined using resonant
  inelastic {X}-{R}ay scattering}.
\newblock \emph{\bibinfo{journal}{Phys. Rev. Lett.}}
  \textbf{\bibinfo{volume}{103}}, \bibinfo{pages}{117003}
  (\bibinfo{year}{2009}).

\bibitem{Haverkort2010}
\bibinfo{author}{Haverkort, M.~W.}
\newblock \bibinfo{title}{Theory of resonant inelastic {X}-{R}ay scattering by
  collective magnetic excitations}.
\newblock \emph{\bibinfo{journal}{Phys. Rev. Lett.}}
  \textbf{\bibinfo{volume}{105}}, \bibinfo{pages}{167404}
  (\bibinfo{year}{2010}).

\bibitem{Fumagalli2019}
\bibinfo{author}{Fumagalli, R.} \emph{et~al.}
\newblock \bibinfo{title}{Polarization-resolved {C}u ${L}_{3}$-edge resonant
  inelastic {X}-{R}ay scattering of orbital and spin excitations in
  {${\mathrm{NdBa}}_{2}{\mathrm{Cu}}_{3}{\mathrm{O}}_{7\ensuremath{-}\ensuremath{\delta}}$}}.
\newblock \emph{\bibinfo{journal}{Phys. Rev. B}} \textbf{\bibinfo{volume}{99}},
  \bibinfo{pages}{134517} (\bibinfo{year}{2019}).

\bibitem{zenodo}
\bibinfo{author}{Zhang, W.} \emph{et~al.}
\newblock \bibinfo{title}{Unravelling the nature of the spin excitations
  disentangled from the charge contributions in a doped cuprate
  superconductor}.
\newblock \emph{\bibinfo{journal}{zenodo.7286412}}  (\bibinfo{year}{2022}).

\bibitem{Bricogne2005}
\bibinfo{author}{Bricogne, G.} \emph{et~al.}
\newblock \bibinfo{title}{{X-ray absorption, refraction and resonant scattering
  tensors in selenated protein crystals: implications for data collection
  strategies in macromolecular crystallography}}.
\newblock \emph{\bibinfo{journal}{J. Appl. Cryst.}}
  \textbf{\bibinfo{volume}{38}}, \bibinfo{pages}{168--182}
  (\bibinfo{year}{2005}).

\bibitem{Nakajima1999}
\bibinfo{author}{Nakajima, R.}, \bibinfo{author}{St\"ohr, J.} \&
  \bibinfo{author}{Idzerda, Y.~U.}
\newblock \bibinfo{title}{Electron-yield saturation effects in ${L}$-edge
  {X}-ray magnetic circular dichroism spectra of {Fe}, {Co}, and {Ni}}.
\newblock \emph{\bibinfo{journal}{Phys. Rev. B}} \textbf{\bibinfo{volume}{59}},
  \bibinfo{pages}{6421--6429} (\bibinfo{year}{1999}).

\bibitem{Henneken2000}
\bibinfo{author}{Henneken, H.}, \bibinfo{author}{Scholze, F.} \&
  \bibinfo{author}{Ulm, G.}
\newblock \bibinfo{title}{Lack of proportionality of total electron yield and
  soft {X}-ray absorption coefficient}.
\newblock \emph{\bibinfo{journal}{J. Appl. Phys.}}
  \textbf{\bibinfo{volume}{87}}, \bibinfo{pages}{257--268}
  (\bibinfo{year}{2000}).

\bibitem{Ruosi2014}
\bibinfo{author}{Ruosi, A.} \emph{et~al.}
\newblock \bibinfo{title}{Electron sampling depth and saturation effects in
  perovskite films investigated by soft {X}-ray absorption spectroscopy}.
\newblock \emph{\bibinfo{journal}{Phys. Rev. B}} \textbf{\bibinfo{volume}{90}},
  \bibinfo{pages}{125120} (\bibinfo{year}{2014}).

\bibitem{Ronay1991}
\bibinfo{author}{Ronay, M.} \emph{et~al.}
\newblock \bibinfo{title}{A new correlation for ${T_\text{c}}$ from {C}u 2p
  absorption}.
\newblock \emph{\bibinfo{journal}{Solid State Communications}}
  \textbf{\bibinfo{volume}{77}}, \bibinfo{pages}{699--702}
  (\bibinfo{year}{1991}).

\bibitem{Nucker1995}
\bibinfo{author}{N\"ucker, N.} \emph{et~al.}
\newblock \bibinfo{title}{Site-specific and doping-dependent electronic
  structure of
  {${\mathrm{YBa}}_{2}$${\mathrm{Cu}}_{3}$${\mathrm{O}}_{\mathit{x}}$} probed
  by {O} 1s and {Cu} 2p {X}-ray-absorption spectroscopy}.
\newblock \emph{\bibinfo{journal}{Phys. Rev. B}} \textbf{\bibinfo{volume}{51}},
  \bibinfo{pages}{8529--8542} (\bibinfo{year}{1995}).

\bibitem{Chabot2010}
\bibinfo{author}{Chabot-Couture, G.} \emph{et~al.}
\newblock \bibinfo{title}{Polarization dependence and symmetry analysis in
  indirect ${K}$-edge {RIXS}}.
\newblock \emph{\bibinfo{journal}{Phys. Rev. B}} \textbf{\bibinfo{volume}{82}},
  \bibinfo{pages}{035113} (\bibinfo{year}{2010}).

\bibitem{Achkar2011}
\bibinfo{author}{Achkar, A.~J.} \emph{et~al.}
\newblock \bibinfo{title}{Bulk sensitive {X}-ray absorption spectroscopy free
  of self-absorption effects}.
\newblock \emph{\bibinfo{journal}{Phys. Rev. B}} \textbf{\bibinfo{volume}{83}},
  \bibinfo{pages}{081106(R)} (\bibinfo{year}{2011}).

\end{thebibliography}

%\clearpage
\setcounter{figure}{0}
\renewcommand{\figurename}{Supplementary Figure}

\begin{figure*}[t]
\centering
\includegraphics[width=0.6\textwidth]{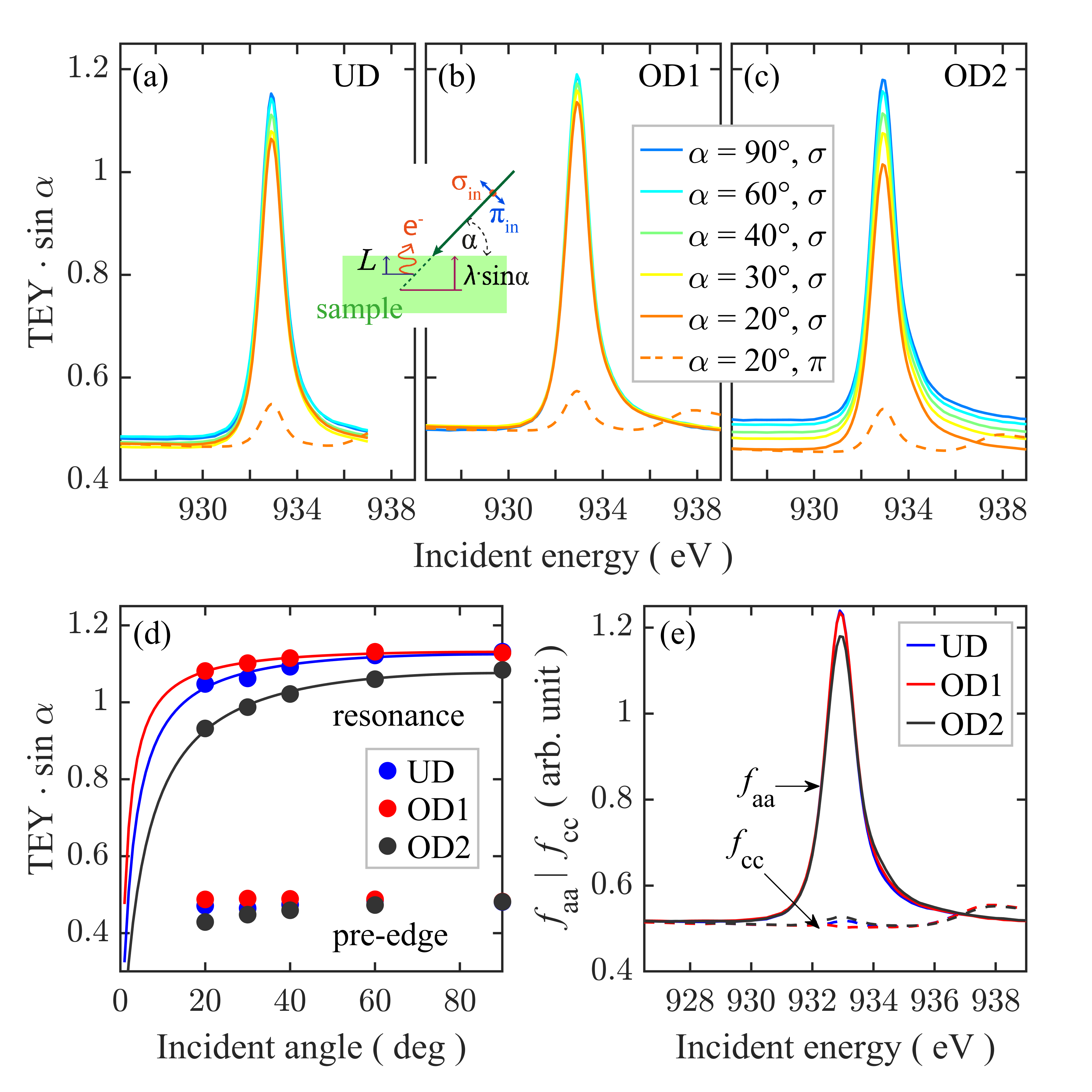}
\caption{{\bf X-ray absorption spectra and the extracted absorption tensor elements $f_{aa}(E)$ and $f_{cc}(E)$ for the three samples.} (a) -(c) The incident-angle dependent XAS of UD, OD1 and OD2 samples, respectively. (d) The intensity of $\text{TEY}\cdot \sin \alpha$ of the resonance peak and pre-edge background as a function of incident angle $\alpha$. (e) The extracted absorption tensor elements $f_{aa}(E)$ and $f_{cc}(E)$ of UD, OD1 and OD2 sample, respectively. Inset: geometry and photon polarizations of the measurements.}\label{XAS}
\end{figure*}

\begin{figure*}[t]
  \includegraphics[width=\textwidth]{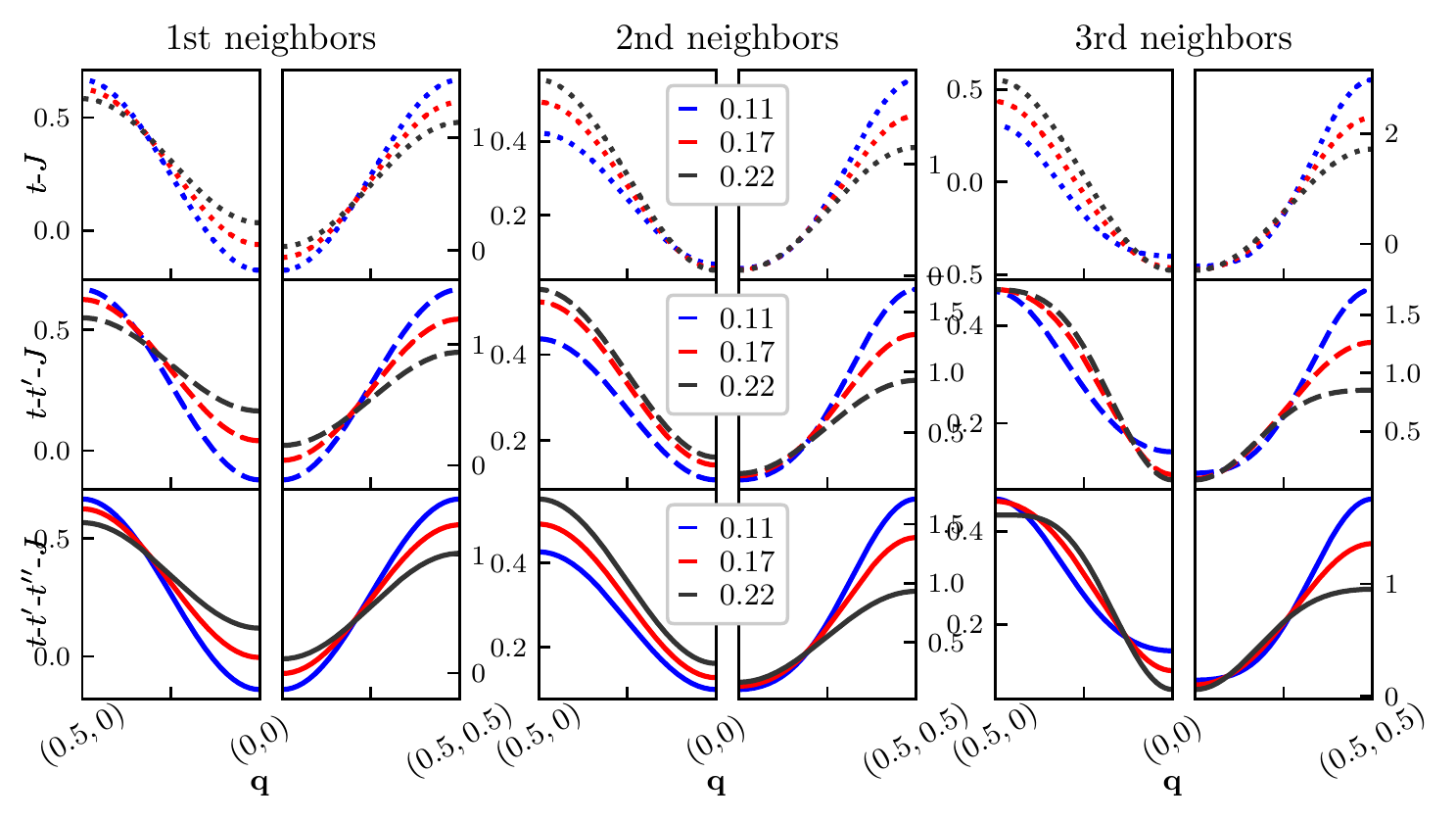}
  \caption{{\bf Detailed contributions of the short-range magnetic correlations and the longer-range electronic hoppings to the theoretical static spin structure factor.} Static spin structure factor $ S(\mathbf q)$ obtained using DMRG on a $6\times 6$ cluster (see text for further details) for the $t$-$J$ (top panels), the $t$-$t'$-$J$ (middle panels), and $t$-$t'$-$t''$-$J$ model (bottom panels). In all cases only a restricted number of short-range spin-spin correlations is nonzero in the Fourier Transform of Eq. 5: solely first neighbors (left panels); solely first and second neighbors (middle panels); and solely first, second, and third neighbors (right panels). Model parameters as in Fig. 5. Note the enlarged momentum coverage w.r.t. Fig. 5 and different scales of $S(\mathbf q)$ for the nodal and anti-nodal directions of the Brillouin zone.}
\end{figure*}

\begin{figure*}[t]
\centering
  \includegraphics[width=0.6\textwidth]{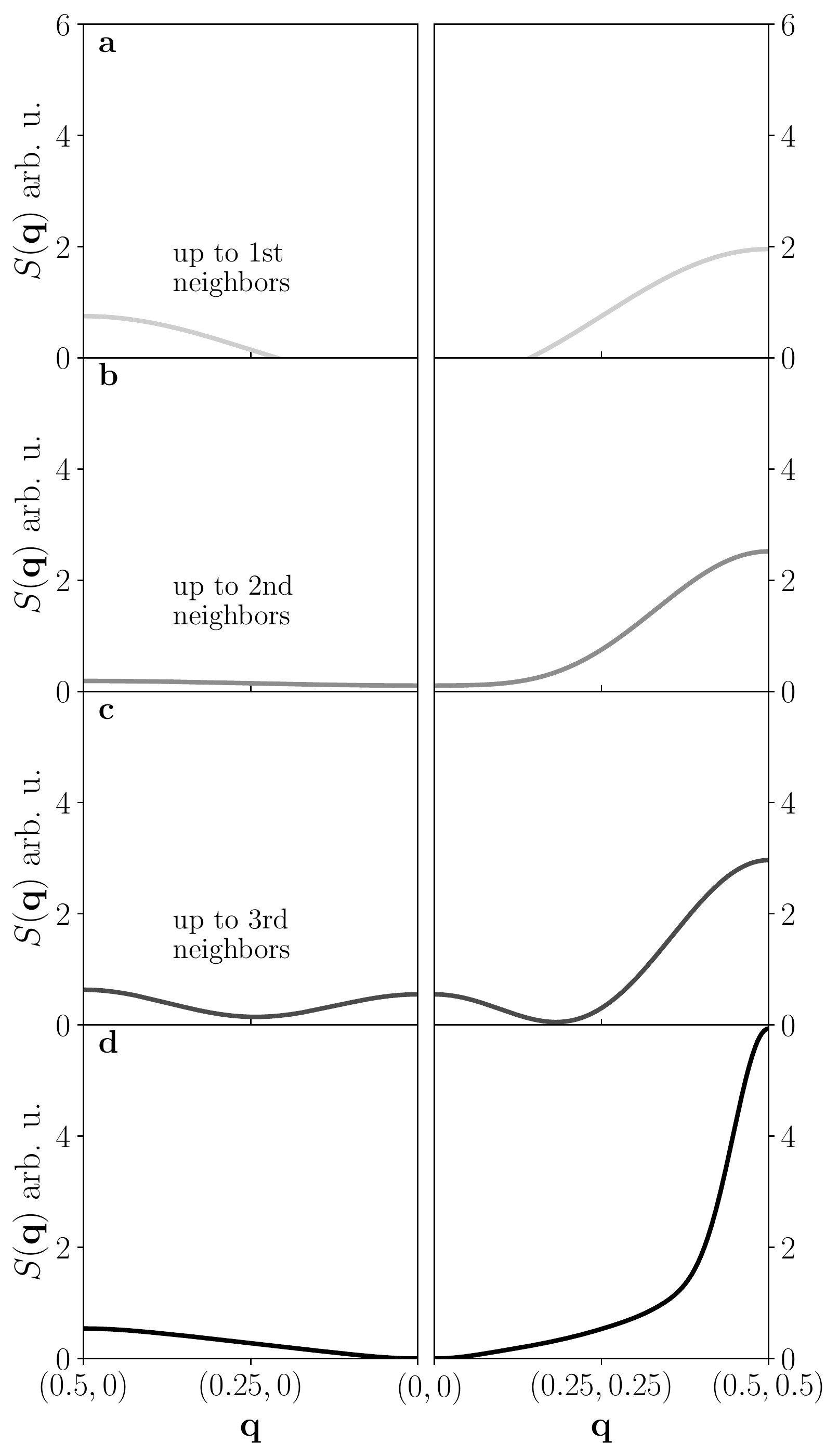}
  \caption{{\bf Theoretical static spin structure factor at half-filling.} Static spin structure factor S(q)  obtained using DMRG on a $6\times 6$ cluster for the half-filled $t$--$J$ model and (as in all other cases) including the edge factor $\lambda$ for a square lattice and keeping only up to first ({\bf a} panel), second ({\bf b} panel) and third ({\bf c} panel) neighbor spin-spin correlations in the Fourier Transform. {\bf d} Static spin structure factor $ S(\mathbf q)$ obtained using DMRG on a $6\times 6$ cluster for the half-filled $t$--$J$ model and (as in all other cases) including the edge factor $\lambda$ for a square lattice.  Note the enlarged momentum coverage w.r.t. Fig. 5 and the same scale in the nodal and anti-nodal direction as well as the agreement with the textbook result of $S(\mathbf q)$ for the Heisenberg model.}
\end{figure*}

\begin{figure*}[t]
\centering
  \includegraphics[width=0.6\textwidth]{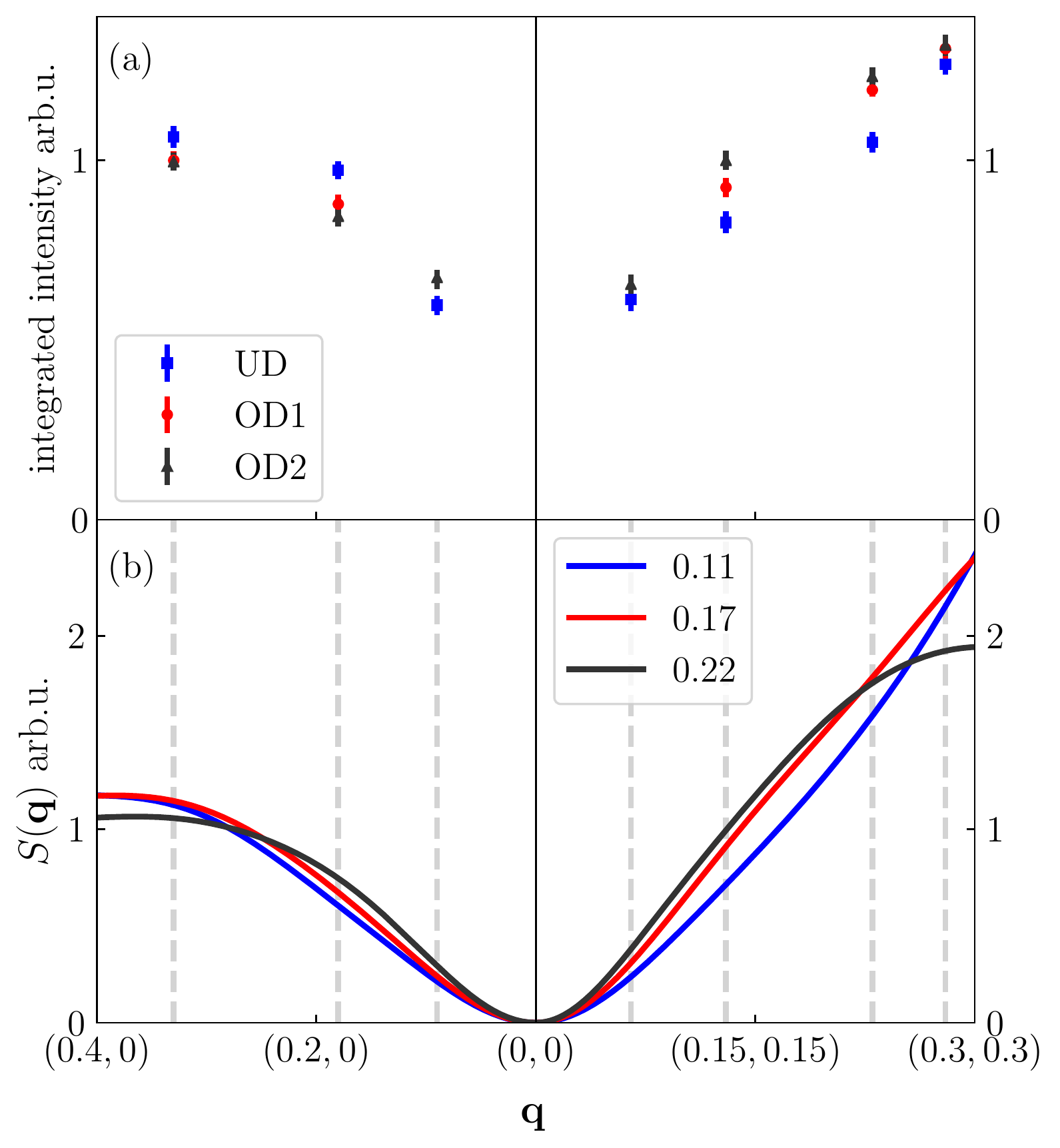}
  \caption{{\bf a} The experimental proxy for the static spin structure factor, i.e. the integrated intensity of the spin excitations as seen by RIXS on Bi2212.  The points are normalised to the value of the intensity at $\mathbf{q}=(0.13,0.13)$ for the OD2 doping level. {\bf b} Static spin structure factor $S(\mathbf{q})$ obtained using DMRG on a $6\times 6$ cluster (see text for further details) for the $t$-$t'$-$J$ model with $t'=-0.5 t$, $J=-0.4 t$. The results are normalised to the value of $S(\mathbf{q})$ at $\mathbf{q}\simeq(0.13,0.13)$ for the doping level $n=0.22$.}
\end{figure*}

\begin{figure*}[t]
\centering
  \includegraphics[width=0.6\textwidth]{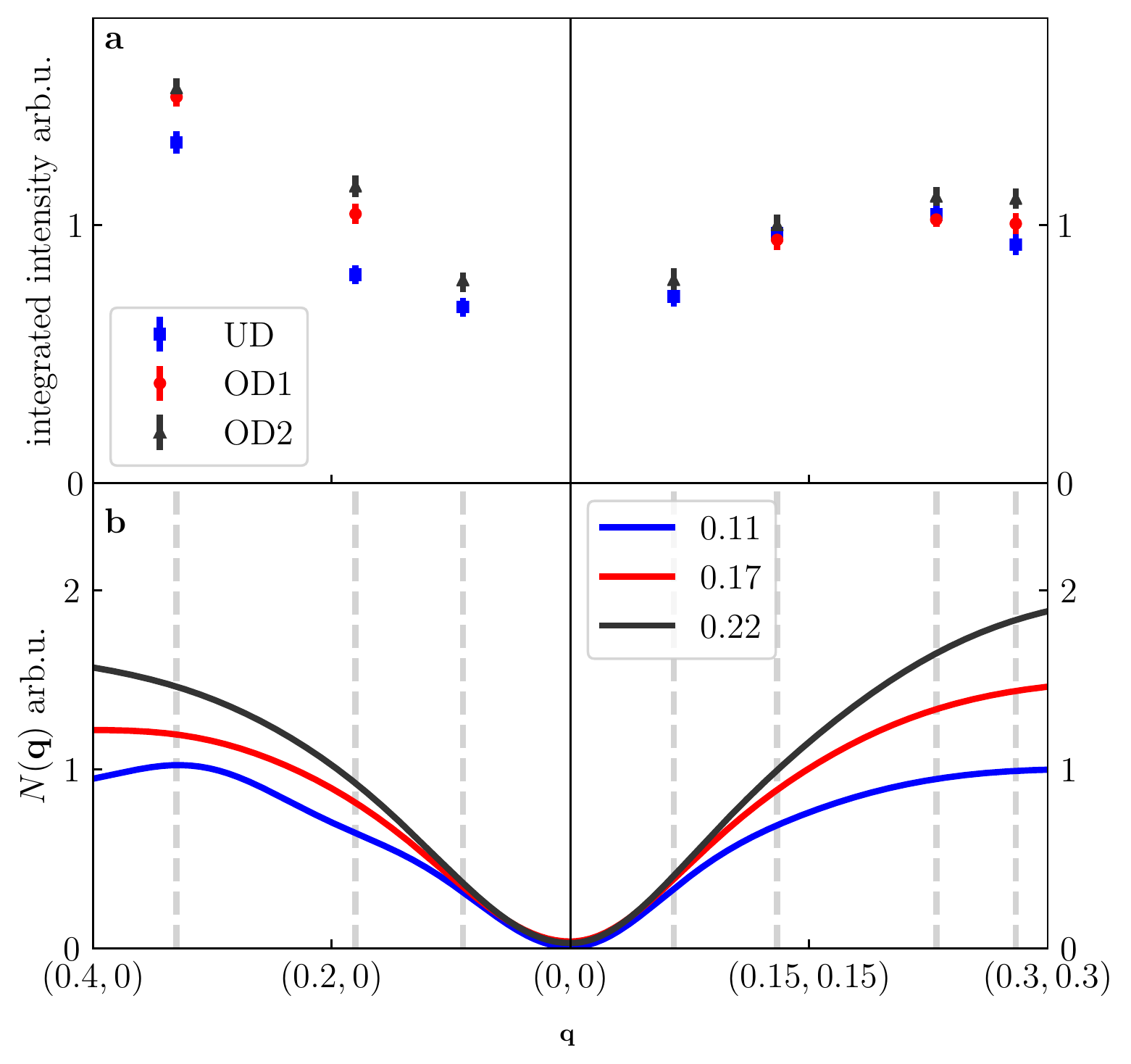}
  \caption{{\bf a} The experimental proxy for the static charge structure factor, i.e. the integrated intensity of the charge excitations as seen by RIXS on Bi2212, see Fig. 3{\bf p} in the main text. The points are normalised to the value of the intensity at $\mathbf{q}=(0.13,0.13)$ for the OD2 doping level. {\bf b} Static charge structure factor $N(\mathbf{q})$ obtained using DMRG on a $6\times 6$ cluster (see text for further details) for the $t$-$t'$-$t''$-$J$ model and three different hole-doping levels. The results are normalised to the value of $N(\mathbf{q})$ at $\mathbf{q}\simeq(0.13,0.13)$ for the doping level $n=0.22$. Model parameters: $J=0.4t$, $t'=-0.3t$, $t''=0.15t$.}
\end{figure*}

\begin{figure*}[t]
    \centering
    \includegraphics[width=0.6\textwidth]{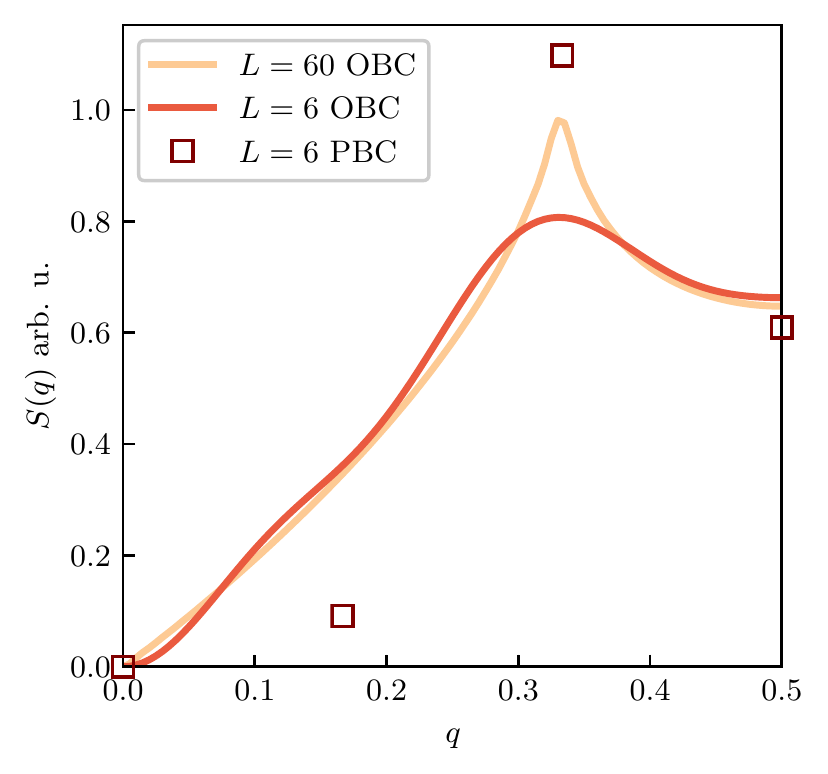}
    \caption{{\bf Usage of OBC vs PBC condition.} We plot results for $S(q)$ for the $t$-$J$ model on a chain with doping level $n=2/3$. Lines show results for OBC on chains of size $L=60$ and $L=6$, edge factors are applied on the smaller cluster. The overall agreement in the evolution of $S(q)$ is incredibily good. Open squares show results at the allowed momenta for a closed chain with $L=6$. Overall, the $S(q)$ values do not compare well with those of the $L=60$ cluster.}
\end{figure*}

\begin{figure*}[t]
\centering
  \includegraphics[width=0.6\textwidth]{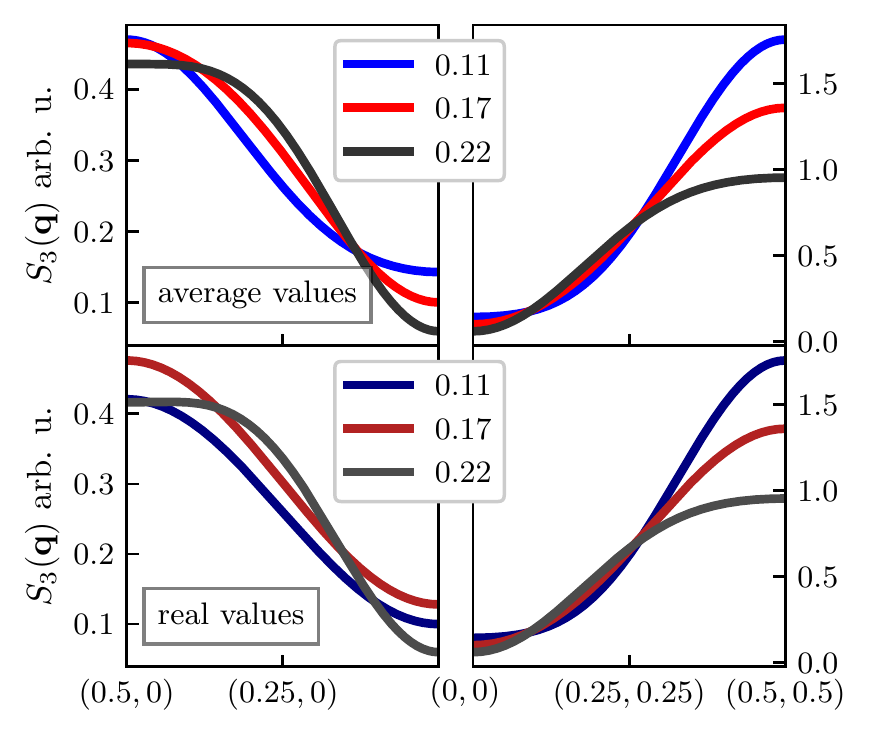}
  \caption{{\bf Theoretical static spin structure factor including averaged and non-averaged short-range magnetic correlations.} Static spin structure factor $ S(\mathbf q)$ obtained using DMRG on a $6\times 6$ cluster for the doped  $t$-$t'$-$t''$-$J$ model with only up to third neighbor spin-spin correlations included in the Fourier transform defining $ S(\mathbf q)$. Top (bottom) panels shows results obtained using an averaged (non-averaged) values of the short magnetic correlations, respectively. Model parameters as in Fig. 5. Note the enlarged momentum coverage w.r.t. Fig. 5 and different scales of $S(\mathbf q)$ for the nodal and anti-nodal directions of the Brillouin zone.}
\end{figure*}

\end{document}